\documentstyle[pra,aps,eqsecnum,multicol]{revtex}
\input psfig.sty

\begin{document}
\title{From Completely Positive Maps to the Quantum Markovian Semigroup
Master Equation}
\author{Daniel A. Lidar,$^1$\thanks{
Current address: Department of Chemistry, 80 St. 
George Street, University of Toronto, Toronto, Ontario M5S 3H6}
Zsolt Bihary,$^2$
and K. Birgitta Whaley$^1$\thanks{
Corresponding author. Tel: 510-643-6820; fax: 510-643-1255;
Email address:whaley\@socrates.berkeley.edu.}}
\address{$^1$Department of Chemistry, University of California, Berkeley,
California 94720 \\
$^2$ Department of Chemistry, University of California, Irvine,
California 92697}

\maketitle

\begin{abstract}
A central problem in the theory of the dynamics of open quantum systems
is the derivation of a rigorous and computationally tractable master equation for the reduced
system density matrix.
Most generally, the evolution of an open quantum system is described by a
completely positive linear map.
We show how to derive a
completely positive Markovian master equation (the Lindblad equation)
from such a map by a coarse graining procedure. We provide a novel and explicit
recipe
for calculating the coefficients of the master equation, using
perturbation theory in the weak-coupling limit. The only parameter
external to our theory is the coarse-graining time-scale. We
illustrate the method by explicitly deriving the master equation for the
spin-boson model. The results are evaluated for the exactly solvable
case of pure dephasing, and an excellent agreement is found within the
timescale where the Markovian approximation is expected to be
valid. The method can be extended in principle to include
non-Markovian effects.
\end{abstract}

\begin{multicols}{2}

\section{Introduction}
The problem of the emergence of irreversible quantum dynamics from
closed-system, unitary dynamics has occupied the attention of many
researchers since the birth of quantum mechanics \cite
{Schrodinger,Wheeler,Davies:76,Alicki:87,Gardiner:book,Weiss:book}. It is
generally believed that an acceptable solution is to view every quantum
system as coupled to an environment, i.e., true quantum systems are always
``open''. The action of the environment is to perform measurements on the
system, thus establishing a preferred ``pointer basis'' and leading to
decoherence \cite{Zurek:81,Giulini:book}. Within a Hamiltonian framework,
recipes for deriving the associated reduced-dynamics were known since the
early nineteen-sixties, starting with the Zwanzig projection technique: one writes
down the Heisenberg equation of motion for the combined system-environment
state, and then projects out the system by tracing out the environment
degrees of freedom \cite{Zwanzig:61}. This yields an integro-differential
equation involving an environment memory kernel, which must be subjected to
approximations in order to become useful. The two main techniques available
are the derivation of master equations by the use of the Born-Markov
approximation \cite{Gardiner:book}, or a representation in terms of path
integrals and influence functionals \cite{Weiss:book}. Unfortunately, in the
former approach it is often unclear whether or not {\em complete positivity}
is preserved in the sequence of approximations one makes \cite{Kohen:97},
while in the latter approach one must make a semi-classical approximation in
order to obtain a tractable theory \cite{Makri:91}.

Complete positivity in reduced dynamics is the very common-sensical
idea\footnote{
See \cite{Pechukas:94,Pechukas+Alicki:95} for a debate concerning this
assertion.} that the open-system dynamics must preserve the positivity of a
system's density matrix (a necessary condition for the probability
interpretation to hold)\ in the presence of any other {\em non-interacting}
system. Building on this notion, two seminal contributions have been made. Kraus
established an ``operator-sum representation'' which describes the most
general completely-positive linear map on the density matrix of a quantum
system \cite{Kraus:83}. This is a formal representation of the dynamics,
which has been used profitably in the quantum information processing
community \cite{Nielsen:book}, but is impractical to use for dynamics
calculations. To address this, Lindblad has derived the most general
completely-positive Markovian semigroup master equation for the dynamics of
the density matrix \cite{Lindblad:76}. This master equation can be integrated
and solved to provide the time-development of the system density matrix.
Both of these results were derived on the basis of axiomatic quantum mechanics.
While systematically satisfying, this approach
nevertheless possesses the disadvantage that the resulting theories are
necessarily phenomenological, in the
sense that they contain no recipe for deriving their parameters from first
principles \cite{Alicki:87}.

Previous formal approaches therefore suffer from one of two disadvantages. 
Either the final equations are not necessarily completely positive, or
they contain 
parameters which are not derived from first principles and must therefore be
treated as phenomenological.
In this paper we provide a derivation of the semigroup master equation (SME)
from the Kraus operator-sum representation (OSR) which overcomes 
both of the above problems.  Thus, the SME we derive is
completely-positive (i.e., it is of Lindblad form), while we can also provide a
recipe
for calculating the parameters that appear in the equation. Our technique
involves a coarse-graining procedure which leaves us with just one
phenomenological parameter: the coarse-graining time-scale $\tau $. The work
presented here is a continuation and generalization of \cite{Bacon:99},
where the derivation of the SME from the OSR was provided for the first
time. Here we verify the validity of the approach in \cite{Bacon:99}, by proving
now that the resulting SME is, as required, completely positive
(Section~\ref{SME}).  Furthermore, we
greatly expand the utility of the derivation by now also showing explicitly how to
calculate the parameters that appear in the SME (Section~\ref{param}).  The
method uses
a perturbative expansion in the system-bath coupling strength. We apply our
formalism to the simple example of a collection of spins coupled to a boson
bath, and compare the result to the exact solution (Section~\ref{example}). We
conclude with an overview and assessment of possible extensions
(Section~\ref{conclusions}).

\section{From the Operator Sum Representation of Reduced Dynamics to the
Semigroup Master Equation}
\label{SME}

\subsection{Brief Review of the Operator Sum Representation}
\label{OSR}

The dynamics of a quantum system $S$ coupled to a bath $B$, which together
form a closed system, evolves unitarily under the combined system-bath
Hamiltonian 
\begin{equation}
{\bf H}_{SB}={\bf H}_{S}{\bf \otimes I}_{B}+{\bf I}_{S}\otimes {\bf
H}_{B}+ 
{\bf H}_{I}.  \label{eq:H}
\end{equation}
Here ${\bf H}_{S}$, ${\bf H}_{B}$ and ${\bf H}_{I}$ are, respectively, the
system, bath and interaction Hamiltonians, and ${\bf I}$ is the identity
operator. Assuming that $S$ and $B$ are initially decoupled, so that the
total initial density matrix is a tensor product of the system and bath
density matrices ($\rho $ and $\rho _{B}$ respectively), the system dynamics
are described by the reduced density matrix:
\begin{equation}
\rho (0)\longmapsto \rho (t)={\rm Tr}_{B}[{\bf U}(\rho \otimes \rho _{B}) 
{\bf U}^{\dagger }].  \label{eq:dyna}
\end{equation}
Here Tr$_{B}$ is the partial trace over the bath and 
\begin{equation}
{\bf U}=\exp (-\frac{i}{\hbar }{\bf H}_{SB}t).  \label{eq:U}
\end{equation}
By using a spectral decomposition for the bath, $\rho _{B}=\sum_{\mu }\wp_{\mu}
|\mu \rangle \langle \mu |$ (where $\sum_{\mu }\wp_{\mu} =1$), and introducing a
basis $\{|n\rangle \}_{n=1}^{N}$ for the $N$-dimensional system Hilbert
space ${\cal H}$, this can be rewritten in the OSR as \cite{Kraus:83}:
\begin{equation}
\rho (t)=\sum_{i=0}^{K}{\bf A}_{i}(t)\,\rho (0)\,{\bf A}_{i}^{\dagger }(t),
\label{eq:OSR}
\end{equation}
where the Kraus operators $\{{\bf A}_{i}\}$ have matrix elements given by 
\cite{Alicki:87}:
\begin{equation}
\lbrack {\bf A}_{i}\rbrack_{mn}(t)=\sqrt{\wp_{\mu} }\langle m|\langle \mu |{\bf U}
(t)|\nu \rangle |n\rangle \;;\qquad i=(\mu ,\nu ).  \label{eq:Amunu}
\end{equation}
$K=N_{B}^{2}$, where $N_{B}$ is the number of bath degrees of freedom.
Also,
by unitarity of ${\bf U}$, one derives the normalization condition
\begin{equation}
\sum_{i=0}^{K}{\bf A}_{i}^{\dagger }{\bf A}_{i}={\bf I}_{S},
\label{eq:OSRnorm}
\end{equation}
which guarantees preservation of the trace of $\rho $: ${\rm Tr}[\rho (t)]= 
{\rm Tr}[\sum_{i}{\bf A}_{i}\,\rho (0)\,{\bf A}_{i}^{\dagger }]={\rm Tr}
[\rho (0)\sum_{i}{\bf A}_{i}^{\dagger }{\bf A}_{i}]={\rm Tr}[\rho (0)].$The
Kraus operators belong to the Hilbert-Schmidt space ${\cal A}({\cal H})$
(itself a Hilbert space) of bounded operators acting on the system Hilbert
space, and are represented by $N\times N$ matrices, just like $\rho $.

\subsection{Fixed-Basis Form of the Operator Sum Representation}
\label{newOSR}

While the OSR evolution equation, Eq.~(\ref{eq:OSR}), is perfectly general,
it presents two major difficulties:\ (i) It is an {\em evolution} equation,
rather than a differential equation, which expresses $\rho (t)$ in terms of
the initial condition and time-dependent operators. Calculating these is
equivalent to diagonalizing the entire system-bath Hamiltonian.  This is
impractical in all but a very few exactly solvable models. (ii) It is not
clear how to separate out the unitary evolution of the system from the
possibly non-unitary one, which occurs from the coupling of the system to
the bath and leads to decoherence. The reason is that in general, each Kraus
operator will contain a contribution from both the unitary and the
non-unitary components of the evolution. When one makes the assumption of
Markovian dynamics, however, as in the semigroup master equation (SME)
[Eqs.~(\ref
{eq:SG1}),(\ref{eq:SG}) below], both of these problems are solved, i.e., one
obtains a differential equation in which there is an explicit separation between
terms leading to unitary and to non-unitary evolution.
This provided the motivation in~\cite{Bacon:99} to develop an alternative
representation of the OSR 
in a form which approaches the form of the
SME as much as is possible, without yet making any Markovian assumption.   We
provide only the main
steps of this derivation here, and refer the interested reader to \cite
{Bacon:99} for full details.

It is convenient for this purpose to introduce a {\em fixed} operator basis
for ${\cal A}({\cal H})$. Let $\{{\bf K}_{\alpha }\}_{\alpha =0}^{M}$, with $
{\bf K}_{0}={\bf I}$, be such a basis, so that the expansion of the Kraus
operators is given by:
\begin{equation}
{\bf A}_{i}(t)=\sum_{\alpha =0}^{M}b_{i\alpha }(t){\bf K}_{\alpha }.
\label{eq:A-F}
\end{equation}
Under this expansion, the OSR evolution equation, Eq.~(\ref{eq:OSR}), becomes
\begin{equation}
\rho (t)=\sum_{\alpha ,\beta =0}^{M}\chi _{\alpha \beta }(t){\bf K}_{\alpha
}\rho (0){\bf K}_{\beta }^{\dagger },  \label{eq:chiOSR}
\end{equation}
where $\chi _{\alpha \beta }(t)$ is the matrix with elements
\begin{equation}
\chi _{\alpha \beta }(t)=\sum_{i=0}^{K}b_{i\alpha }(t)b_{i\beta }^{\ast }(t).
\label{eq:defchi}
\end{equation}
The matrix $\chi $ is clearly Hermitian, with positive diagonal elements.
With some algebraic manipulation \cite{Bacon:99} one can transform Eq.~(\ref
{eq:chiOSR}) into:
\begin{eqnarray}
\frac{\partial \rho (t)}{\partial t} &=&-{\frac{i}{\hbar }}[\dot{{\bf S}}
(t),\rho (0)] +{\frac{1}{2}}\sum_{\alpha ,\beta =1}^{M}\dot{\chi}_{\alpha
\beta }(t) \times  \nonumber \\
&&\left( [{\bf K}_{\alpha },\rho (0){\bf K}_{\beta }^{\dagger }]+[{\bf K}
_{\alpha }\rho (0),{\bf K}_{\beta }^{\dagger }]\right) .  \label{eq:newOSR2}
\end{eqnarray}
where ${\bf S}(t)$ is the hermitian operator defined by
\begin{equation}
{\bf S}(t)=\frac{i\hbar }{2}\sum_{\alpha =1}^{M}\left[ \chi _{\alpha 0}(t) 
{\bf K}_{\alpha }-\chi _{0\alpha }(t){\bf K}_{\alpha }^{\dagger }\right] .
\label{eq:defJ}
\end{equation}
Eq.~(\ref{eq:newOSR2}) is the desired result: it represents a fixed-basis
OSR evolution equation, with a strong resemblance to the SME, as we now
detail.

\subsection{From the Fixed-Basis Operator Sum Representation
Equation to the Semigroup Master Equation}
\label{OSR-SG}

\subsubsection{Derivation of the Semigroup Master Equation by a Coarse
Graining Procedure}

We recall that in the semigroup approach, under the assumptions of (i)
Markovian dynamics, (ii) initial decoupling between the system and the bath,
and (iii) the requirement of complete positivity, the system evolves
according to the SME \cite{Alicki:87,Lindblad:76}:
\begin{eqnarray}
\frac{\partial \rho (t)}{\partial t} &=&{\tt L}[\rho (t)]\equiv -\frac{i}{
\hbar }[{\bf H}_{S},\rho (t)]+{\tt L}_{D}[\rho (t)]  \label{eq:SG1} \\
{\tt L}_{D}[\rho (t)] &=&\frac{1}{2}\sum_{\alpha ,\beta =1}^{M}a_{\alpha
\beta }([{\bf F}_{\alpha },\rho (t){\bf F}_{\beta }^{\dagger }]+[{\bf F}
_{\alpha }\rho (t),{\bf F}_{\beta }^{\dagger }]),  \label{eq:SG}
\end{eqnarray}
where $a_{\alpha \beta }$ is a constant positive semi-definite matrix. This
equation bears a clear resemblance to Eq.~(\ref{eq:newOSR2}). Analyzing the
differences between the SME Eq.~(\ref{eq:SG}) and this OSR evolution
equation (\ref{eq:newOSR2}) allows one to understand the precise manner in
which the semigroup evolution arises from the OSR evolution under the above
mentioned three conditions. An important difference between these two
equations is the fact that the SME provides a prescription for determining $
\rho (t)$ at all times $t$, given $\rho (t^{\prime })$ as an initial
condition at any other time $
t> t^{\prime }\geq 0$, whereas Eq.~(\ref{eq:newOSR2}) determines $\rho (t)$ in
terms of $\rho (0),$ i.e., at the special time $t=0$ where the system and
the bath are in a product state.

In \cite{Bacon:99} a coarse-graining procedure was introduced which
allows to transform the exact Eq.~(\ref{eq:newOSR2}) to the
approximate SME. For 
convenience we repeat and clarify the derivation here. We consider three
time-scales:\ (i)\ the inverse of the bath density of states
frequency-cutoff $\tau _{c}$, (ii) a coarse-graining time-scale $\tau $ which is
essentially the time-scale for the bath's ``memory'' to disappear (the
definition will be made more precise below), and (iii) a system
time-scale $\theta  
$ which is the typical time-scale for changes in the system density-matrix
in the frame rotating with the system Hamiltonian. We require that 
\begin{equation}
\tau _{c}\ll \tau \ll \theta ,  \label{eq:taus}
\end{equation}
and course-grain the evolution of the system in terms of $\tau $: $\rho
_{j}=\rho (j\tau )$; $\chi _{\alpha \beta ;j}=\chi _{\alpha \beta }(j\tau )$
, $j$ an integer. Further, rewriting the OSR Eq.~(\ref{eq:newOSR2}) as $\rho
(t)={\bf \Lambda }(t)\rho (0)$ and defining $\tilde{{\tt L}}(t)$ through $ 
{\bf \Lambda }(t)={\rm T}\exp \left[ \int_{0}^{t}\tilde{{\tt L}}(s)ds\right] 
$ ({\rm T} indicates time-ordering)\ we have
\begin{equation}
{\frac{\partial \rho (t)}{\partial t}}=\tilde{{\tt L}}(t)[\rho (t)].
\label{eq:OSRtrue}
\end{equation}
Define $\tilde{{\tt L}}_{j}=\int_{j\tau }^{(j+1)\tau }\tilde{{\tt L}}(s)ds$, with $\tau n=t$: $\int_{0}^{t}\tilde{{\tt L}}(s)ds=\tau \sum_{j=0}^{n-1} 
\tilde{{\tt L}}_{j}$. Next we make the assumption that on the
coarse-graining time-scale $\tau $, the evolution generators $\tilde{{\tt L}}
(t)$ commute in the ``average'' sense that $\left[ \tilde{{\tt L}}_{j}, 
\tilde{{\tt L}}_{k}\right] =0,\forall j,k$. Physically, we imagine this
operation as arising from the ``resetting'' of the bath density operator
over the time-scale $\tau $. {\em This means that }$\tau ${\em \ must be
larger than any characteristic bath time-scale}, and explains the
requirement $\tau _{c}\ll \tau $. Under this assumption, the evolution of
the system is Markovian when $t\gg \tau $: ${\bf \Lambda }
(t)=\prod_{j=0}^{n-1}\exp \left[ \tau \tilde{{\tt L}}_{j}\right] .$ Further,
under the discretization of the evolution, this product form of the
evolution implies that $\rho _{j+1}=\exp \left[ \tau \tilde{{\tt L}}_{j} 
\right] [\rho _{j}]$. In the limit of $\tau \ll t$ we expand this
exponential, to find that
\begin{equation}
{\frac{\rho _{j+1}-\rho _{j}}{\tau }}=\tilde{{\tt L}}_{j}[\rho _{j}].
\label{eq:rhodot}
\end{equation}
This equation is simply a discretization of Eq.~(\ref{eq:OSRtrue}) under the
assumption that $\tau \ll \theta $, where $\theta $ is the time-scale of
change for the system density matrix. Notice in particular that the RHS of
Eq.~(\ref{eq:rhodot}) contains the {\em average} value of $\tilde{{\tt L}}
(t) $ over the interval. Now, from the OSR evolution equation (\ref
{eq:newOSR2}), we know the explicit form of $\tilde{{\tt L}}(t)$ over the
first interval from $0$ to $\tau $. Discretizing over this interval we find
that
\begin{eqnarray}
{\frac{\rho _{1}-\rho _{0}}{\tau }}& =&-{\frac{i}{\hbar }}\left[
\left\langle \dot{{\bf S}}\right\rangle ,\rho _{0}\right]  \nonumber \\
&+&{\frac{1}{2}}\sum_{\alpha ,\beta =1}^{M}\left\langle \dot{\chi}_{\alpha
\beta }\right\rangle \left( [{\bf K}_{\alpha },\rho _{0}{\bf K}_{\beta
}^{\dagger }]+[{\bf K}_{\alpha }\rho (0),{\bf K}_{\beta }^{\dagger }]\right)
\nonumber \\
&\equiv& \tilde{{\tt L}}_{0}[\rho _{0}],
\end{eqnarray}
where
\begin{equation}
\left\langle X\right\rangle \equiv {\frac{1}{\tau }}\int_{0}^{\tau }X(s)ds.
\label{eq:t-ave}
\end{equation}
Thus, in the sense of the course graining above we have arrived at an
explicit form for $\tilde{{\tt L}}_{0}$. However, deriving an explicit form
for $\tilde{{\tt L}}_{1}$ and for higher terms beyond this first interval is
impossible because Eq.~(\ref{eq:newOSR2}) gives the evolution in terms of $
\rho (0)$. Since we have made the assumption that the bath ``resets'' over
the time-scale $\tau $, we expect the bath to interact with the system in
the same manner over every $\tau $-length coarse-grained interval. This is
equivalent to assuming that $\tilde{{\tt L}}_{i}=\tilde{{\tt L}}_{0},\forall
i$ (which of course is the most trivial way of satisfying the Markovian
evolution condition $[\tilde{{\tt L}}_{i},\tilde{{\tt L}}_{j}]=0,\forall i,j$
). Then, under the natural identification of the ${\bf K}$'s with the ${\bf 
F }$'s of the SME, and using Eq.~(\ref{eq:rhodot}), one is led to the well
known form of the semigroup equation of motion:
\begin{eqnarray}
\frac{\partial \rho (t)}{\partial t} &=&-\frac{i}{\hbar }[\left\langle {\dot{
{\bf S}}}\right\rangle ,\rho (t)]  \nonumber \\
&&+{\frac{1}{2}}\sum_{\alpha ,\beta =1}^{M}\left\langle {\dot{\chi}}_{\alpha
\beta }\right\rangle \left( [{\bf K}_{\alpha },\rho (t){\bf K}_{\beta
}^{\dagger }]+[{\bf K}_{\alpha }\rho (t),{\bf K}_{\beta }^{\dagger }]\right)
\nonumber \\
&&  \label{eq:newOSR3}
\end{eqnarray}

\subsubsection{Positivity of the Coefficient Matrix}

The positive semi-definiteness of the coefficient matrix $a_{\alpha \beta }$
in {Eq.~(\ref{eq:SG}) is a sufficient condition for the preservation of
complete positivity of the system dynamics \cite{Alicki:87}. Thus, to
complete the identification of Eq.~(\ref{eq:newOSR3}) as a Lindblad
equation, it only remains to be shown that }$\left\langle {\dot{\chi}}
_{\alpha \beta }\right\rangle $ is positive semi-definite. To do so let us
show first that $\chi _{\alpha \beta }$ itself is positive semi-definite,
i.e., that for any vector ${\bf c}$, the matrix $\chi $ satisfies ${\bf c}
\chi {\bf c}^{\ast t}\geq 0$: 
\begin{eqnarray}
{\bf c}\chi {\bf c}^{\ast t} &=&\sum_{\alpha \beta }c_{\alpha }^{\ast }\chi
_{\alpha \beta }c_{\beta }=\sum_{i,\alpha \beta }c_{\alpha }^{\ast
}b_{i\alpha }b_{i\beta }^{\ast }c_{\beta }  \nonumber \\
&=&\sum_{i}\left| \sum_{\alpha }c_{\alpha }^{\ast }b_{i\alpha }\right|
^{2}\geq 0,  \label{eq:chipos}
\end{eqnarray}
where we used Eq.~(\ref{eq:defchi}). Next, 
\begin{equation}
\left\langle {\dot{\chi}}_{\alpha \beta }\right\rangle ={\frac{1}{\tau }}
\int_{0}^{\tau }{\dot{\chi}}_{\alpha \beta }dt={\frac{1}{\tau }}\left( {\chi 
}_{\alpha \beta }(\tau )-{\chi }_{\alpha \beta }(0)\right) ,
\label{eq:chicheck}
\end{equation}
so that we must show that ${\chi }(0)$ does not spoil the positivity. Now,
from Eqs.{~(\ref{eq:U}) and (\ref{eq:Amunu})} ${\bf A}_{i}(0)=\sqrt{\wp_{\mu} }
\langle \mu |{\bf U}(0)|\nu \rangle =\sqrt{\wp_{\mu} }\delta _{\mu \nu }{\bf I}
_{S}=\sum_{\alpha =0}^{M}b_{i\alpha }(0){\bf K}_{\alpha }$, so that $
b_{i\alpha }(0)=\sqrt{\wp_{\mu} }\delta _{\alpha 0}\delta _{i,(\mu ,\mu )}$
(recall ${\bf K}_{0}={\bf I}_{S}$). Thus 
\begin{eqnarray}
{\chi }_{\alpha \beta }(0) &=&\sum_{i}b_{i\alpha }(0)b_{i\beta }^{\ast
}(0)=\sum_{i=(\mu ,\nu )}\wp_{\mu} \delta _{\alpha 0}\delta _{\beta 0}\delta
_{i,(\mu ,\mu )}  \nonumber \\
&=&\delta _{\alpha 0}\delta _{\beta 0}.  \label{eq:chiat0}
\end{eqnarray}
But in Eq.~(\ref{eq:newOSR3}) we are concerned with $\left\langle {\dot{\chi}
}_{\alpha \beta }\right\rangle $ only for $\alpha ,\beta \geq 1$, so that
finally, from {Eq.~(\ref{eq:chicheck}), the submatrix }$\left\langle {\dot{
\chi}}_{\alpha \beta }\right\rangle $ with $\alpha ,\beta \geq 1$ is indeed
positive semi-definite. The important conclusion is that Eq.~(\ref
{eq:newOSR3}) is in Lindblad form, i.e., it preserves complete positivity.
This establishes the validity of our result for the SME, and should be
contrasted with projection-operator type derivations of the master equation 
\cite{Gardiner:book,Pollard:96}, which do not necessarily satisfy the
complete positivity criterion.

\subsubsection{Separating Out the Hamiltonian}

We can write {Eq.~(\ref{eq:newOSR3})} in an alternative form which
distinguishes between the system and bath contributions to the unitary part
of the evolution. Because Eq.~(\ref{eq:newOSR2}) is linear in the $\chi
_{\alpha \beta }(t)$ matrix, one can calculate $\chi _{\alpha \beta
}^{(0)}(t)$ for the isolated system and hence define the new terms which
come about from the coupling of the system to the bath: $\chi _{\alpha \beta
}(t)=\chi _{\alpha \beta }^{(0)}(t)+\chi _{\alpha \beta }^{(1)}(t)$. The
terms which correspond to the isolated system will then produce a normal $
-(i/\hbar )[{\bf H}_{S},\rho (t)]$ Liouville term in Eq.~(\ref{eq:newOSR3}). Thus Eq.~(\ref{eq:newOSR3}) can be rewritten as
\begin{eqnarray}
\frac{\partial \rho (t)}{\partial t} &=&-{\frac{i}{\hbar }}\left[ {\bf H}
_{S}+\left\langle {\dot{{\bf S}}}^{(1)}\right\rangle ,\rho (t)\right] 
\nonumber \\
&&+{\ \frac{1}{2}}\sum_{\alpha ,\beta =1}^{M}\left\langle {\dot{\chi}}
_{\alpha \beta }\right\rangle \left( [{\bf K}_{\alpha },\rho (t){\bf K}
_{\beta }^{\dagger }]+[{\bf K}_{\alpha }\rho (t),{\bf K}_{\beta }^{\dagger
}]\right) .  \nonumber \\
&&  \label{eq:newOSR4}
\end{eqnarray}
Identifying $\left\langle {\dot{\chi}}_{\alpha \beta
}\right\rangle $ with $a_{\alpha \beta }$, and ${\bf K}_{\alpha }$ with $ 
{\bf F}_{\alpha }$, this is seen to be equivalent to Eqs.({\ref{eq:SG1})-(\ref{eq:SG}),
except for the presence of the second term in the Liouvillian. This second
term $\left\langle {\dot{{\bf S}}}^{(1)}\right\rangle $, inducing unitary
dynamics on the system, is referred to as the {\em Lamb shift}. It
explicitly describes the effect the bath has on the unitary part of the
system dynamics, and ``renormalizes'' the system Hamiltonian. It is often
implicitly assumed to be present in Eq.~(\ref{eq:SG1}) \cite{Beck:93}. }

In summary, we have shown in this Section how coarse-graining the evolution
over the bath memory time-scale $\tau $ allows one to understand the
connection between the OSR evolution and the semigroup dynamics. The
importance of Eq.~(\ref{eq:newOSR2}) lies in the fact that it allows one to
pinpoint the exact point at which the assumption of Markovian dynamics is
made.  Furthermore, due to the general likeness of its form to the SME, it
provides an easily translatable connection from the non-Markovian
OSR to the Markovian SME. Notice also that the assumption of Markovian
dynamics introduces an arrow of time in the evolution of the system, through
the ordering of the environmental states: The system evolves through time in
the direction of successive resettings of the bath. Additionally, it is important to
note that we have
shown that our procedure leads to an explicitly Lindbladian (completely
positive) form of the SME, as written in final form in Eq.~(\ref{eq:newOSR4}).

Finally, we address the question of the inclusion of non-Markovian effects.
The approach presented here also offers a route to a systematic
inclusion of non-Markovian effects, i.e., higher order dynamics which
include bath
memory terms.  Such a derivation of a ``post-Markovian'' master
equation is
a long
sought-after goal of the field of open quantum systems. Several
attempts
have been reported, but generally the resulting equations are not
satisfactory because complete positivity is violated
\cite{Yu:2000}. In
the context of the present approach, the formal extension to go beyond
the
Markovian
regime can be made by replacing the assumption that the evolution
generators
$\tilde{{\tt L}}_j$ commute to first order [see text below
Eq.~(\ref{eq:OSRtrue})], by a higher order commutator. The
derivation of this commutator and the resulting post-Markovian master
equation is left to a future publication.

\section{Explicit Derivation of the Semigroup Master Equation Parameters}
\label{param}

We can now exploit the coarse-grained first-order (in time)
perturbation expansion of the OSR, Eq.~(\ref{eq:newOSR2}), made in the
previous section, in order to
derive the {\em explicit} form of the parameters and operators appearing in
the resulting SME. To do so, it turns out to be most convenient to work in the
interaction picture (IP) defined with respect to the free system and bath
Hamiltonians. Let us the system-bath
interaction Hamiltonian of Eq.~(\ref{eq:H}) in the following
perfectly general form:\footnote{Note that $\{{\bf S}_{\alpha }\}$ and
$\{{\bf B}_{\alpha }\}$ are not assumed to be linear operators, and
that any interaction Hamiltonian can be decomposed into a
sum of terms acting separately on system and bath. Furthermore, we
allow $\{{\bf S}_{\alpha }\}$ and $\{{\bf B}_{\alpha }\}$ to be
time-dependent.}
\begin{equation}
{\bf H}_{I}=\sum_{\alpha }\lambda _{\alpha }{\bf S}_{\alpha }\otimes {\bf B}
_{\alpha },  \label{eq:H_I}
\end{equation}
where $\{{\bf S}_{\alpha }\}$ and $\{{\bf B}_{\alpha }\}$ are the system and
bath operators respectively, and $\{\lambda _{\alpha }\}$ are coupling
coefficients. In the IP we do not have to deal directly with the free
system and bath Hamiltonians. However, as will be seen below, we do recover
the Lamb shift.

\subsection{The Interaction Picture}

Transformation to the IP is accomplished by means of the unitary
operator 
\begin{equation}
{\bf U}_{T}=\exp \left( -it{\bf H}_{S}\right) \otimes \exp \left( -it{\bf H}
_{B}\right) \equiv {\bf U}_{S}\otimes {\bf U}_{B}.
\end{equation}
Operators in the IP will be denoted using explicit time dependence (and
where there already was a time dependence, with an $I$ subscript). Thus: 
\begin{equation}
{\bf H}_{I}(t)={\bf U}_{T}^{\dagger }{\bf H}_{I}{\bf U}_{T}=\sum_{\alpha
}\lambda _{\alpha }{\bf S}_{\alpha }(t)\otimes {\bf B}_{\alpha }(t)
\end{equation}
where 
\begin{eqnarray}
{\bf S}_{\alpha }(t) &=&{\bf U}_{S}^{\dagger }{\bf S}_{\alpha }{\bf U}
_{S}=\sum_{\beta }p_{\alpha \beta }(t){\bf S}_{\beta }  \label{eq:Sexp} \\
{\bf B}_{\alpha }(t) &=&{\bf U}_{B}^{\dagger }{\bf B}_{\alpha }{\bf U}
_{B}=\sum_{\beta }q_{\alpha \beta }(t){\bf B}_{\beta },  \label{eq:Bexp}
\end{eqnarray}
with $p_{\alpha \beta }(0)=q_{\alpha \beta }(0)=\delta _{\alpha \beta }$.
The density matrix for the system and bath combined is denoted $\rho _{{\rm 
tot}}(t)$ in the Schr\"{o}dinger picture, and is transformed to the IP by $
\rho _{{\rm tot,}I}(t)={\bf U}_{T}^{\dagger }\rho _{{\rm tot}}(t){\bf
U}_{T}$. The dynamics of $\rho _{{\rm tot,}I}(t)$ is governed by the unitary
propagator ${\bf U}(t)={\bf U}_{T}^{\dagger }\exp (-it{\bf H}_{SB}){\bf U}
_{T}$, where ${\bf H}_{SB}$ is the full system-bath Hamiltonian: $\rho _{
{\rm tot,}I}(t)={\bf U}(t)\rho _{{\rm tot,}I}(0){\bf U}^{\dagger }(t)$. The
Schr\"{o}dinger and interaction pictures coincide at $t=0$ so that $\rho _{
{\rm tot,}I}(0)=$ $\rho _{{\rm tot}}(0)=\rho (0)\otimes \rho _{B}(0)$. It is
a standard exercise to show that \cite{March:book} 
\begin{eqnarray}
{\bf U}(t) &=&{\rm T}\exp \left[ -\frac{i}{\hbar }\int_{0}^{t}{\bf H}
_{I}(\tau )d\tau \right]  \nonumber \\
&=&{\bf I}+\sum_{n=1}^{\infty }\frac{\left( -i\right) ^{n}}{n!}{\bf U}_{n}(t)
\label{eq:Un}
\end{eqnarray}
where 
\begin{equation}
{\bf U}_{n}(t)\equiv
\int_{0}^{t}dt_{n}\int_{0}^{t}dt_{n-1}...\int_{0}^{t}dt_{1}{\rm T}\left\{ 
{\bf H}_{I}(t_{1})\cdots {\bf H}_{I}(t_{n})\right\} .
\end{equation}
The Dyson time-ordered product is defined with respect to any set of
operators ${\bf O}_{i}(t_{\tau })$ as \cite{March:book}:\ ${\rm T}\left\{ 
{\bf O}_{1}(t_{1})\cdots {\bf O}_{n}(t_{n})\right\} ={\bf O}_{\tau
_{1}}(t_{\tau _{1}})\cdots {\bf O}_{\tau _{n}}(t_{\tau _{n}})$, where $
t_{\tau _{1}}>t_{\tau _{2}}>...>t_{\tau _{n}}$. The system density matrix in
the IP is obtained, as usual, by tracing over the bath, which leads to the
OSR: 
\begin{equation}
\rho _{I}(t)={\rm Tr}_{B}\left[ \rho _{{\rm tot,}I}(t)\right]
=\sum_{i=0}^{K} {\bf A}_{i}(t)\,\rho (0)\,{\bf A}_{i}^{\dagger }(t),
\end{equation}
where the Kraus operators are now defined in the IP: 
\begin{equation}
{\bf A}_{i}(t)=\sqrt{\wp_{\mu} }\langle \mu |{\bf U}(t)|\nu \rangle .
\label{eq:AiIP}
\end{equation}
Repeating the derivation of Sections \ref{newOSR}, \ref{OSR-SG} we thus
obtain the very same form for the SME as in Eq.~(\ref{eq:newOSR3}), but now
it is a SME for the interaction representation, $\rho _{I}(t)$. Finally, the
transformation back to the Schr\"{o}dinger picture is accomplished by: 
\begin{equation}
\rho (t)={\bf U}_{S}\rho _{I}(t){\bf U}_{S}^{\dagger }.  \label{eq:ItoS}
\end{equation}

\subsection{Perturbation Theory Expansion of the Kraus Operators}

Our next task is to calculate the Kraus operators. We do so by using the
expansion for ${\bf U}(t)$ and Eqs.~(\ref{eq:Un}),(\ref{eq:AiIP}). We have
then: 
\begin{equation}
{\bf A}_{i}(t)=\sqrt{\wp_{\mu} }\delta _{\mu \nu }{\bf I}_{S}+\sum_{n=1}^{\infty } 
{\bf A}_{i}^{(n)}(t),  \label{eq:Aiexp}
\end{equation}
where 
\begin{eqnarray}
{\bf A}_{i}^{(n)}(t) &=&\sqrt{\wp_{\mu} }\frac{\left( -i\right) ^{n}}{n!}
\sum_{\alpha _{1},...,\alpha
_{n}}\int_{0}^{t}dt_{n}\int_{0}^{t}dt_{n-1}...\int_{0}^{t}dt_{1}\times 
\nonumber \\
&&{\rm T}\left[ \prod_{j=1}^{n}\lambda _{\alpha _{j}}{\bf S}_{\alpha
_{j}}(t_{j})\right] \langle \mu |{\rm T}\left[ \prod_{j=1}^{n}{\bf B}
_{\alpha _{j}}(t_{j})\right] |\nu \rangle ,  \nonumber \\
&&
\end{eqnarray}
and we used $\left[ {\bf S}_{\alpha _{i}}(t_{i}),{\bf B}_{\alpha
_{j}}(t_{j}) \right] =0$ to separate the time-ordering operations. ${\bf A}
_{i}^{(n)}$ is proportional to $\left( \lambda _{\alpha }\right) ^{n}$, so
that in the weak-coupling case of $\lambda _{\alpha }\ll 1$, we can truncate
the expansion at small $n$.

\subsection{First Order Case}
\label{firstorder}

First we note that from Eq.~(\ref{eq:Aiexp}), with ${\bf K}_{0}={\bf I}_{S}$
: $b_{i0}(t)=\sqrt{\wp_{\mu} }\delta _{\mu \nu }$. Now, let us calculate the
expression for $n=1$. In this case there is no need to worry about time
ordering, and we have: 
\begin{eqnarray}
{\bf A}_{i=\mu \nu }^{(1)}(t) &=&-i\sqrt{\wp_{\mu} }\sum_{\beta
}\int_{0}^{t}dt_{1}\lambda _{\beta }{\bf S}_{\beta }(t_{1})\langle \mu |{\bf 
B}_{\beta }(t_{1})|\nu \rangle  \nonumber \\
&=&-it\sqrt{\wp_{\mu} }\sum_{\alpha \beta \gamma }{\bf S}_{\alpha }\lambda _{\beta
}\langle \mu |{\bf B}_{\gamma }|\nu \rangle \Gamma _{\beta }^{\alpha \gamma
}(t)  \nonumber \\
&\equiv &\sum_{\alpha }b_{i\alpha }(t){\bf K}_{\alpha },
\end{eqnarray}
where the second line follows using Eqs.~(\ref{eq:Sexp}),(\ref{eq:Bexp}) the
third from the fixed-basis operator expansion in Eq.~(\ref{eq:A-F}), and we
defined 
\begin{equation}
\Gamma _{\alpha }^{\beta \gamma }(t)\equiv \frac{1}{t}\int_{0}^{t}dt_{1}p_{
\alpha \beta }(t_{1})q_{\alpha \gamma }(t_{1}).  \label{eq:Gamma}
\end{equation}
This dimensionless quantity thus depends entirely on the transformation to
the interaction picture, i.e., it contains no information on the system-bath
coupling, but only on the internal system and bath dynamics. Next, let us
identify ${\bf K}_{\alpha }={\bf S}_{\alpha }$ (the system operators)\ and
assume that our basis is trace-orthogonal: 
\begin{equation}
{\rm Tr}\left[ {\bf S}_{\alpha }^{\dagger }{\bf S}_{\beta }\right] =\delta
_{\alpha \beta }/N_{\alpha },  \label{eq:TrSaSb}
\end{equation}
where $N_{\alpha }$ is a normalization constant. Then 
\begin{equation}
b_{i\alpha }(t)=-it\sqrt{\wp_{\mu} }\sum_{\alpha ^{\prime }\alpha ^{\prime \prime
}}\lambda _{\alpha ^{\prime }}\langle \mu |{\bf B}_{\alpha ^{\prime \prime
}}|\nu \rangle \Gamma _{\alpha ^{\prime }}^{\alpha \alpha ^{\prime \prime
}}(t)\qquad \alpha \geq 1.  \label{eq:b_ialpha}
\end{equation}
Using these results and $\chi _{\alpha \beta }(t)=\sum_{i=\mu \nu
}b_{i\alpha }(t)b_{i\beta }^{\ast }(t)$ we can reconstruct the $\chi $
matrix: $\chi _{00}(t)=\sum_{\mu }\wp_{\mu} =1$, and for $\alpha \geq 1$: 
\begin{eqnarray}
\chi _{\alpha 0}(t) &=&\sum_{i=\mu \nu }b_{i\alpha }(t)b_{i0}^{\ast
}(t)=\sum_{\mu }\sqrt{\wp_{\mu} }b_{\mu \mu ,\alpha }(t)  \nonumber \\
&=&-it\sum_{\alpha ^{\prime }\alpha ^{\prime \prime }}\lambda _{\alpha
^{\prime }}\langle {\bf B}_{\alpha ^{\prime \prime }}\rangle _{B}\Gamma
_{\alpha ^{\prime }}^{\alpha \alpha ^{\prime \prime }}(t)  \label{eq:chia0}
\end{eqnarray}
where we used 
\begin{equation}
\langle {\bf X}\rangle _{B}\equiv {\rm Tr}[\rho _{B}{\bf X}]=\sum_{\mu }
\wp_{\mu} \langle \mu |{\bf X}|\mu \rangle ,
\end{equation}
defining the bath-averaged expectation value of an arbitrary operator ${\bf 
X }$. Finally, for both $\alpha ,\beta \geq 1$ 
\begin{eqnarray}
\chi _{\alpha \beta }(t) &=&\sum_{i=\mu \nu }b_{i\alpha }(t)b_{i\beta
}^{\ast }(t)  \nonumber \\
&=&t^{2}\sum_{\alpha ^{\prime }\alpha ^{\prime \prime }\beta ^{\prime }\beta
^{\prime \prime }}\lambda _{\alpha ^{\prime }}\lambda _{\beta ^{\prime
}}^{\ast }\langle {\bf B}_{\alpha ^{\prime \prime }}{\bf B}_{\beta ^{\prime
\prime }}^{\dagger }\rangle _{B}\Gamma _{\alpha ^{\prime }}^{\alpha \alpha
^{\prime \prime }}(t)\left( \Gamma _{\beta ^{\prime }}^{\beta \beta ^{\prime
\prime }}(t)\right) ^{\ast }  \nonumber \\
&&  \label{eq:chiab}
\end{eqnarray}
where in the last line we used the completeness relation $\sum_{\nu }|\nu
\rangle \langle \nu |={\bf I}_{B}$.

Now, as shown in Eqs.~(\ref{eq:chicheck}),(\ref{eq:chiat0}): 
\begin{eqnarray}
a_{\alpha \beta } &=&\left\langle {\dot{\chi}}_{\alpha \beta }\right\rangle
= {\ \ \frac{1}{\tau }}\left( {\chi }_{\alpha \beta }(\tau )-{\chi }_{\alpha
\beta }(0)\right)  \nonumber \\
&=&\frac{{\chi }_{\alpha \beta }(\tau )}{\tau },  \label{eq:<chi>}
\end{eqnarray}
unless both $\alpha =\beta =0$ (in which case $\left\langle {\dot{\chi}}
_{00}\right\rangle =0$). Together with the SME\ Eq.~(\ref{eq:newOSR3}) we
thus have all the ingredients. In particular, we can calculate the $
\left\langle {\dot{{\bf S}}}\right\rangle $ Lamb shift term in Eq.~(\ref
{eq:newOSR3}) from Eq.~(\ref{eq:defJ}): 
\begin{eqnarray}
\left\langle {\dot{{\bf S}}}\right\rangle &=&\frac{i}{2}\sum_{\alpha
}\langle {\dot{\chi}}_{\alpha 0}\rangle {\bf S}_{\alpha }-\langle {\dot{\chi}
}_{\alpha 0}\rangle ^{\ast }{\bf S}_{\alpha }^{\dagger }  \nonumber \\
&=&\frac{1}{2}\sum_{\alpha }\phi _{\alpha }{\bf S}_{\alpha }+\phi _{\alpha
}^{\ast }{\bf S}_{\alpha }^{\dagger },  \label{eq:Sdot}
\end{eqnarray}
where 
\begin{equation}
\phi _{\alpha }\equiv \sum_{\alpha ^{\prime }\alpha ^{\prime \prime
}}\lambda _{\alpha ^{\prime }}\langle {\bf B}_{\alpha ^{\prime \prime
}}\rangle _{B}\Gamma _{\alpha ^{\prime }}^{\alpha \alpha ^{\prime \prime
}}(\tau )  \label{eq:phi_alpha}
\end{equation}
is a correlation function which contributes to the Lamb shift. Note that
unlike in Eq.~(\ref{eq:newOSR4}), $\left\langle {\dot{{\bf S}}}\right\rangle 
$ does not contain the system Hamiltonian, as indeed it should not in the IP.

As for the decoherence term, we find 
\begin{eqnarray}
a_{\alpha \beta } &=&\left\langle {\dot{\chi}}_{\alpha \beta }\right\rangle 
\nonumber \\
&=&\tau \sum_{\alpha ^{\prime }\alpha ^{\prime \prime }\beta ^{\prime }\beta
^{\prime \prime }}\lambda _{\alpha ^{\prime }}\lambda _{\beta ^{\prime
}}^{\ast }\langle {\bf B}_{\alpha ^{\prime \prime }}{\bf B}_{\beta ^{\prime
\prime }}^{\dagger }\rangle _{B}\Gamma _{\alpha ^{\prime }}^{\alpha \alpha
^{\prime \prime }}(\tau )\left( \Gamma _{\beta ^{\prime }}^{\beta \beta
^{\prime \prime }}(\tau )\right) ^{\ast }.  \nonumber \\
&&  \label{eq:chidot}
\end{eqnarray}
Note that both the Lamb shift parameters $\phi _{\alpha }$ and the
decoherence parameters $a_{\alpha \beta }$ formally depend on the
coarse-graining time $\tau $. Since $\tau $ is a ``dummy'' differentiation
parameter in our theory, the dependence upon it should disappear in an
explicit calculation. We deal with this in the example studied in Section~ 
\ref{example}.

\subsection{Second Order Case}
\label{2ndorder}

Expanding the Kraus operators to second order yields:
\begin{eqnarray}
{\bf A}_{i=\mu \nu }^{(2)}(t) &=&-\frac{\sqrt{\wp_{\mu} }}{2}\sum_{\alpha
_{1},\alpha _{2}}\lambda _{\alpha _{1}}\lambda _{\alpha
_{2}}\int_{0}^{t}dt_{2}\int_{0}^{t}dt_{1}\times  \nonumber \\
&&{\rm T}\left[ {\bf S}_{\alpha _{1}}(t_{1}){\bf S}_{\alpha _{2}}(t_{2}) 
\right] \langle \mu |{\rm T}\left[ {\bf B}_{\alpha _{1}}(t_{1}){\bf B}
_{\alpha _{2}}(t_{2})\right] |\nu \rangle ,  \nonumber \\
&=&-\frac{\sqrt{\wp_{\mu} }t^{2}}{2}\sum_{\alpha _{1}\alpha _{2}}{\bf S}_{\alpha
_{1}}{\bf S}_{\alpha _{2}}\times  \nonumber \\
&&\sum_{\beta _{1}\beta _{2},\gamma _{1}\gamma _{2}}\lambda _{\beta
_{1}}\lambda _{\beta _{2}}\langle \mu |{\bf B}_{\gamma _{1}}{\bf
B}_{\gamma
_{2}}|\nu \rangle \Gamma _{\beta _{1}\beta _{2}}^{\alpha _{1}\alpha
_{2};\gamma _{1}\gamma _{2}}(t)  \label{eq:A2order}
\end{eqnarray}
where 
\begin{eqnarray}
\Gamma _{\alpha _{1}\alpha _{2}}^{\beta _{1}\beta _{2};\gamma _{1}\gamma
_{2}}(t)&\equiv& \frac{1}{t^{2}}\int_{0}^{t}dt_{2}\int_{0}^{t}dt_{1}{\rm T} 
\left[ p_{\alpha _{1}\beta _{1}}(t_{1})p_{\alpha _{2}\beta _{2}}(t_{2}) 
\right] \times \nonumber \\
&{\rm T}&\left[ q_{\alpha _{1}\gamma _{1}}(t_{1})q_{\alpha _{2}\gamma
_{2}}(t_{2})\right] .
\end{eqnarray}
We need to compare this expression to the expansion of the Kraus operators
in terms of the fixed basis, ${\bf A}_{i}(t)=\sum_{\alpha =0}^{M}b_{i\alpha
}(t){\bf K}_{\alpha }$ [Eq.~(\ref{eq:A-F})]. To do so we must now extend the
fixed basis set so that it includes product terms: 
\begin{equation}
\sum_{\alpha =0}^{M}b_{i\alpha }(t){\bf K}_{\alpha }=\sum_{\alpha
_{1}=0}b_{i;\alpha _{1}}(t){\bf S}_{\alpha _{1}}+\sum_{\alpha _{1},\alpha
_{2}=0}b_{i;\alpha _{1}\alpha _{2}}(t){\bf S}_{\alpha _{1}}{\bf S}_{\alpha
_{2}}. 
\end{equation}
Comparing this expansion to Eq.~(\ref{eq:A2order}) we can read off the
second order time-dependent coefficients $b_{i;\alpha _{1}\alpha _{2}}$ as 
\begin{equation}
b_{i;\alpha _{1}\alpha _{2}}=-\frac{\sqrt{\wp_{\mu} }t^{2}}{2}\sum_{\beta
_{1}\beta _{2},\gamma _{1}\gamma _{2}}\lambda _{\beta _{1}}\lambda
_{\beta
_{2}}\langle \mu |{\bf B}_{\gamma _{1}}{\bf B}_{\gamma _{2}}|\nu \rangle
\Gamma _{\beta _{1}\beta _{2}}^{\alpha _{1}\alpha _{2};\gamma _{1}\gamma
_{2}}(t), 
\end{equation}
provided that the basis of system operators is closed under multiplication
(or more generally:\ is trace-orthonormal for all products of basis
elements). The decoherence and Lamb shift parameters can then be calculated
as in the first-order case.

\subsection{Putting It All Together}

In conclusion, we have derived an explicit form for the SME of Eq.~(\ref
{eq:newOSR3}) (with ${\bf S}_{\alpha }$ replacing ${\bf K}_{\alpha }$). To
find the full SME for a given problem, it is necessary to:

\begin{enumerate}
\item Identify the system operators $\{{\bf S}_{\alpha }\}$ in the
interaction Hamiltonian (and recall that these operators must be
trace-orthogonal in our formalism).

\item Solve for the time-dependent system and bath operators in the
interaction picture [Eqs.~(\ref{eq:Sexp}),(\ref{eq:Bexp})], and thus find $
\Gamma $ from Eq.~(\ref{eq:Gamma}).

\item Calculate the expectation values of the bath operators. The results of
this step will depend on the initial state of the bath (e.g., thermal
equilibrium, coherent state, squeezed state, etc.).

\item Use the results of the previous steps to calculate the Lamb shift term 
$\left\langle {\dot{{\bf S}}}\right\rangle $, the decoherence matrix $
\left\langle {\dot{\chi}}_{\alpha \beta }\right\rangle $, and finally, to write
down the SME.
\end{enumerate}

Since this SME is of the Lindblad form \cite{Lindblad:76} it is guaranteed
to preserve positivity of the density matrix. Systematic corrections may be
derived by continuing the expansion in Eq.~(\ref{eq:Aiexp}) to higher orders
in $n$.

\section{Example: Spin-Boson Model}
\label{example}

\subsection{Pure Dephasing of Multiple Qubits}

As a concrete and simple example of the procedure described above, we
consider the form of the SME derived for a collection of independent
two-level systems (qubits) coupled via a phase-damping (non-dissipative)
interaction to a boson bath. The Hamiltonians in this spin-boson model are 
\cite{Leggett:87}:
\begin{eqnarray}
{\bf H}_{S} &=&-\frac{1}{2}\sum_{i}\hbar \omega _{0}^{i}{\sigma
}_{z}^{i},
\label{eq:H_S} \\
{\bf H}_{B} &=&\sum_{k}\hbar \omega _{k}\left( {\bf N}_{k}+\frac{1}{2}
\right) ,  \label{eq:H_B} \\
{\bf H}_{I} &=&\sum_{i,k}{\sigma }_{z}^{i}\otimes \left( \lambda _{k}^{i} 
{\bf b}_{k}+\lambda _{k}^{i\ast }{\bf b}_{k}^{\dagger }\right) ,
\label{eq:H_Isb}
\end{eqnarray}
where $\hbar \omega _{0}^{i}$ are the qubit energies, $\lambda _{k}^{i}$ are
coupling coefficients, ${\bf b}_{k}$ and ${\bf b}_{k}^{\dagger }$ are the $
k^{{\rm th}}$ bath mode annihilation and creation operators obeying the
boson commutation relation $\left[ {\bf b}_{k},{\bf b}_{l}^{\dagger }\right]
={\bf 1}\delta _{kl}$, and ${\bf N}_{k}={\bf b}_{k}^{\dagger }{\bf b}_{k}$
is the number operator. Comparing to Eq.${~}$(\ref{eq:H_I}) we read off the
system operators as ${\bf S}_{\alpha }={\sigma }_{z}^{i}$, and the bath
operators as ${\bf B}_{\alpha }={\bf b}_{k}$. Below we deal with the
required modifications to our treatment of the indices in order to account
for these assignments.

We assume that the boson bath is in thermal equilibrium at temperature $
T=1/\left( k_{B}\beta \right) $ ($k_{B}$ is the Boltzman constant). Thus the
bath density matrix is $\rho _{B}=\frac{1}{Z}e^{-\beta {\bf
H}_{B}}=\frac{1}{
Z}\sum_{\mu }e^{-\beta E_{\mu }}|\mu \rangle \langle \mu |,$where $\mu
=\{n_{1},n_{2},...,n_{k},...\}$ are the numbers of quanta in all bath modes, 
$E_{\mu }$ is the energy of the field at a given occupation $\mu $, and $
Z(T)={\rm Tr}[\exp (-\beta {\bf H}_{B})]=\prod_{k}e^{-\beta \hbar \omega
_{k}/2}/\left( 1-e^{-\beta \hbar \omega _{k}}\right) $ is the canonical
partition function. Some useful results for the average number of quanta in
the $k^{{\rm th}}$ bath mode and the averages of the creation and
annihilation operators are:
\begin{eqnarray}
\langle {\bf b}_{k}^{\dagger }{\bf b}_{l}\rangle _{B} &=&\langle {\bf b}_{k} 
{\bf b}_{l}^{\dagger }\rangle -\delta _{k,l}=\delta _{k,l}\frac{1}{e^{\beta
\hbar \omega _{k}}-1},  \nonumber \\
\langle {\bf b}_{k}^{\dagger }\rangle _{B} &=&\langle {\bf b}_{k}\rangle
_{B}=\langle {\bf b}_{k}^{\dagger }{\bf b}_{l}^{\dagger }\rangle
_{B}=\langle {\bf b}_{k}{\bf b}_{l}\rangle _{B}=0.  \label{eq:<b>}
\end{eqnarray}
We now proceed to calculate the various quantities appearing in the SME.

\subsubsection{Form of the Interaction Representation Operators}

Formally, we need to solve Eqs.~(\ref{eq:Sexp}),(\ref{eq:Bexp}) for the
time-dependent system and bath operators. In the present simple example,
however it is clear that since $\sigma _{z}^{i}$ commutes with the system
Hamiltonian, ${\sigma }_{z}^{i}(t)={\sigma }_{z}^{i}(0)\equiv {\sigma }
_{z}^{i}$. Further, it is an elementary exercise to show that ${\bf b}
_{k}(t)={\bf b}_{k}e^{i\omega _{k}t}$. Therefore the interaction Hamiltonian
in the IP is: 
\begin{equation}
{\bf H}_{I}(t)=\sum_{i,k}{\sigma }_{z}^{i}\otimes (\lambda
_{k}^{i}e^{i\omega _{k}t}{\bf b}_{k}+\lambda _{k}^{i\ast }e^{-i\omega
_{k}t} 
{\bf b}_{k}^{\dagger }).  \label{eq:H_I(t)}
\end{equation}

\subsubsection{Calculation of the Lamb Shift and Decoherence Parameters}

In the derivation of Section \ref{firstorder} the interaction Hamiltonian
was expressed as a sum over the single index $\alpha $, in order to reduce
the clutter of indices to a minimum. However, as seen from the interaction
Hamiltonian of Eq.~(\ref{eq:H_Isb}), in reality we need more indices. In
particular, we need to clarify the indexation of $\Gamma _{\alpha }^{\beta
\gamma }$of Eq.~(\ref{eq:Gamma}): each of the indices $\alpha $,$\beta $ and 
$\gamma $ may now correspond to a qubit position (denoted $i$ or $j$), a
Pauli matrix index (denoted $\xi =x,y,z$), or a bath mode (denoted $k$
or $l$). Qubits variables can have both position and Pauli indices,
but bath 
variables have only a mode index. We will use a comma to separate qubit and
bath variables, as in $\alpha =\left( i\xi ,k\right) $, when all three
indices are needed. When one of the indices is irrelevant it will simply be
dropped. To separate groups of indices, such as the $\beta \gamma $ in $
\Gamma _{\alpha }^{\beta \gamma }$, we will use a semicolon. For example, $
\Gamma _{iz,k}^{j\xi ;l}$ is short for $\Gamma _{\alpha }^{\beta \gamma }$
with $\alpha =\left( iz,k\right) $, $\beta =(j\xi ,k^{\prime })$, $\gamma
=(j^{\prime }\xi _{1},l)$, where $k^{\prime }$, $j^{\prime }$ and $\xi _{1}$
are irrelevant. With these preparations we are now ready to calculate $
\Gamma _{\alpha }^{\beta \gamma }$. By comparing ${\sigma }_{z}^{i}(t)={\
\sigma }_{z}^{i}$, ${\bf b}_{k}(t)={\bf b}_{k}e^{i\omega _{k}t}$ to Eqs.~( 
\ref{eq:Sexp}),(\ref{eq:Bexp}) and using the correct index convention, we
have that $p_{z\xi }^{ii^{\prime }}(t)=\delta _{ii^{\prime }}\delta _{\xi z}$
and $q_{kk^{\prime \prime }}(t)=\delta _{kk^{\prime \prime }}e^{i\omega
_{k}t}$. Therefore: 
\begin{eqnarray}
\Gamma _{iz,k}^{i^{\prime }\xi ;k^{\prime \prime }}(t) &=&\frac{1}{t}
\int_{0}^{t}dt_{1}p_{z\xi }^{ii^{\prime }}(t_{1})q_{kk^{\prime \prime
}}(t_{1})  \nonumber \\
&=&\delta _{ii^{\prime }}\delta _{kk^{\prime \prime }}\delta _{\xi
z}e^{i\omega _{k}t/2}{\rm sinc}\frac{\omega _{k}t}{2} 
\label{eq:GammaNMR}
\end{eqnarray}
where ${\rm sinc}\left( x\right) \equiv \sin \left( x\right) /x$.

Before proceeding to calculate the Lamb shift and decoherence parameters, we
should note that in the definition of ${\bf S}_{\alpha }$ and ${\bf B}
_{\alpha }$ in Eq.~(\ref{eq:H_I}), each ${\bf S}_{\alpha }$ is coupled to a $
{\bf B}_{\alpha }$ with the same index $\alpha $, whereas in the present
case each ${\bf S}_{\alpha }$ (i.e., $\sigma _{z}$) is coupled to both ${\bf 
B}_{\alpha }$ (i.e., ${\bf b}_{k}$)\ and ${\bf B}_{\alpha }^{\dagger }$. Let
us briefly again suppress for clarity the $ij,kl$ indices of the present
example. By linearity, the required modification is clearly that Eq.~(\ref
{eq:b_ialpha}) should be replaced with 
\begin{eqnarray}
b_{\mu \nu ,\alpha }(t) &=&-it\sqrt{\wp_{\mu} }\sum_{\alpha ^{\prime }\alpha
^{\prime \prime }}[\lambda _{\alpha ^{\prime }}\langle \mu |{\bf B}_{\alpha
^{\prime \prime }}|\nu \rangle \Gamma _{\alpha ^{\prime }}^{\alpha \alpha
^{\prime \prime }}(t)  \nonumber \\
&+&\lambda _{\alpha ^{\prime }}^{\ast }\langle \mu |{\bf B}_{\alpha ^{\prime
\prime }}^{\dagger }|\nu \rangle \left( \Gamma _{\alpha ^{\prime }}^{\alpha
\alpha ^{\prime \prime }}(t)\right) ^{\ast }]\qquad \alpha \geq 1
\end{eqnarray}
(where it is assumed that the expansion coefficients of ${\bf S}_{\alpha }$,
the $p_{\alpha \beta }(t)$, are real, as in the example treated here). Using
Eqs.~(\ref{eq:chia0}),(\ref{eq:phi_alpha}) this leads to Lamb shift
parameters of the form: 
\begin{equation}
\phi _{\alpha }=2\sum_{\alpha ^{\prime }\alpha ^{\prime \prime }}{\rm Re} 
\left[ \lambda _{\alpha ^{\prime }}\langle {\bf B}_{\alpha ^{\prime \prime
}}\rangle _{B}\Gamma _{\alpha ^{\prime }}^{\alpha \alpha ^{\prime \prime
}}(\tau )\right] =0,
\end{equation}
since using Eq.~(\ref{eq:<b>}) expectation values of creation and
annihilation operators between number states vanish. As for the decoherence
part, using Eq.~(\ref{eq:<b>}) again we find that $\langle {\bf B}_{\alpha
^{\prime \prime }}{\bf B}_{\beta ^{\prime \prime }}\rangle _{B}$ and $
\langle {\bf B}_{\alpha ^{\prime \prime }}^{\dagger }{\bf B}_{\beta ^{\prime
\prime }}^{\dagger }\rangle _{B}$ vanish, so that using Eqs.~(\ref{eq:chiab}
),(\ref{eq:chidot}), the decoherence parameters are: 
\begin{eqnarray}
\left\langle {\dot{\chi}}_{\alpha \beta }\right\rangle &=&\tau \sum_{\alpha
^{\prime }\alpha ^{\prime \prime }\beta ^{\prime }\beta ^{\prime \prime
}}\lambda _{\alpha ^{\prime }}\lambda _{\beta ^{\prime }}^{\ast }\left[
\langle {\bf B}_{\alpha ^{\prime \prime }}{\bf B}_{\beta ^{\prime \prime
}}^{\dagger }\rangle _{B}\Gamma _{\alpha ^{\prime }}^{\alpha \alpha ^{\prime
\prime }}(\tau )\Gamma _{\beta ^{\prime }}^{\beta \beta ^{\prime \prime
}}(\tau )^{\ast }\right.  \nonumber \\
&&\left. +\langle {\bf B}_{\alpha ^{\prime \prime }}^{\dagger }{\bf B}
_{\beta ^{\prime \prime }}\rangle _{B}\Gamma _{\alpha ^{\prime }}^{\alpha
\alpha ^{\prime \prime }}(\tau )^{\ast }\Gamma _{\beta ^{\prime }}^{\beta
\beta ^{\prime \prime }}(\tau )\right] ,  \label{eq:chiNMR}
\end{eqnarray}
or, using the results for the spin-boson case: 
\begin{eqnarray}
\left\langle {\dot{\chi}}_{i\xi ,j\xi _{1}}\right\rangle &=&\tau
\sum_{i^{\prime }j^{\prime },k^{\prime }l^{\prime };k^{\prime \prime
}l^{\prime \prime }}\lambda _{k^{\prime }}^{i^{\prime }}\lambda _{l^{\prime
}}^{j^{\prime }\ast }\times  \nonumber \\
&&\left[ \langle {\bf b}_{k^{\prime \prime }}{\bf b}_{l^{\prime \prime
}}^{\dagger }\rangle _{B}\Gamma _{i^{\prime }z,k^{\prime }}^{i\xi
;k^{\prime
\prime }}(\tau )\left( \Gamma _{j^{\prime }z,l^{\prime }}^{j\xi
_{1};l^{\prime \prime }}(\tau )\right) ^{\ast }\right.  \nonumber \\
&&\left. +\langle {\bf b}_{k^{\prime \prime }}^{\dagger }{\bf b}_{l^{\prime
\prime }}\rangle _{B}\left( \Gamma _{i^{\prime }z,k^{\prime }}^{i\xi
;k^{\prime \prime }}(\tau )\right) ^{\ast }\Gamma _{j^{\prime }z,l^{\prime
}}^{j\xi _{1};l^{\prime \prime }}(\tau )\right]  \nonumber \\
&=&\tau \delta _{\xi z}\delta _{\xi _{1}z}\sum_{k}\lambda _{k}^{i}\lambda
_{k}^{j\ast }{\rm sinc}^{2}\left( \omega _{k}\tau /2\right) \coth \frac{
\beta \hbar \omega _{k}}{2}  \nonumber \\
&=&a_{ij}^{zz}.
\end{eqnarray}
Our final result for the SME in the interaction picture\ can thus be written
as 
\begin{equation}
\frac{\partial \rho _{I}(t)}{\partial t}={\ \frac{1}{2}}
\sum_{i,j}a_{ij}^{zz}\left( [{\sigma }_{z}^{i},\rho _{I}(t){\sigma }
_{z}^{j}]+[{\sigma }_{z}^{i}\rho _{I}(t),{\sigma }_{z}^{j}]\right) ,
\end{equation}
where 
\begin{equation}
a_{ij}^{zz}=\frac{\tau }{\hbar ^{2}}\sum_{k}\lambda _{k}^{i}\lambda
_{k}^{j\ast }{\rm sinc}^{2}\left( \omega _{k}\tau /2\right) \coth \frac{
\beta \hbar \omega _{k}}{2}.  \label{eq:aijSB}
\end{equation}
and we reintroduced $\hbar $.

As commented above, the dependence on $\tau $ should disappear after all is
done.
Let us see how this comes about within a simple continuum model. If we assume
that $|\lambda _{k}|^{2}$ only depends on $\omega _{k}$, we can rewrite this
expression as an integral over $\omega $:
\begin{eqnarray}
a_{ij}(\tau ) &=&\frac{1}{\hbar ^{2}}\int_{0}^{\infty }d\omega \,g(\omega)\lambda ^{i}(\omega )\lambda ^{j\ast }(\omega )\times  \nonumber \\
&&\tau {\rm sinc}^{2}\frac{\omega \tau }{2}\coth \frac{\beta \hbar \omega }{
2 },
\end{eqnarray}
where $g(\omega )$ is the density of states. In any reasonable physical
model the density of states has a cut-off frequency $\omega _{c}$. Now, $
\tau {\rm sinc}^{2}\left( \omega \tau /2\right) $ is a function that peaks
strongly at $0$, has a width $1/\tau $ and its integral is finite: $\frac{1}{
\pi }\int_{0}^{\infty }d\omega \,\tau {\rm sinc}^{2}\left( \omega \tau
/2\right) =1$, i.e., 
\begin{eqnarray}
\tilde{\delta}(\tau ,\omega ) &\equiv &\frac{1}{\pi }\tau {\rm sinc}
^{2}\left( \omega \tau /2\right)  \nonumber \\
\lim_{\tau =\infty }\tilde{\delta}(\tau ,\omega ) &=&\delta (\omega )
\label{eq:deltafunc}
\end{eqnarray}
But $\tau $, the coarse-graining timescale, must indeed be large compared to
the timescale of the bath $\tau _{c}=1/\omega _{c}$ [recall Eq.~(\ref
{eq:taus})], so in this limit we can perform the integral and we finally get:
\begin{equation}
a_{ij}^{zz}=\frac{\pi }{\hbar ^{2}}\lim_{\omega =0}g(\omega )\lambda
^{i}(\omega )\lambda ^{j}(\omega )^{\ast }\coth \frac{\beta \hbar \omega }{2}
.
\end{equation}
Thus, the dependence on $\tau $ has indeed disappeared.

We now apply this result to the case of a single two-level atom coupled to
a harmonic bath. In the case of phonons and electromagnetic radiation, the
interaction couples to the amplitude of the oscillators: ${\bf x}=\sqrt{
\hbar /\left( 2m\omega \right) }({\bf b}^{\dagger }+{\bf b})$, so that $
|\lambda (\omega )|^{2}\propto 1/\omega $. At the relevant low-frequency
regime we can equivalently use the high-temperature result: 
\begin{equation}
\langle {\bf b}^{\dagger }{\bf b}\rangle _{B}=\langle {\bf bb}^{\dagger
}\rangle _{B}\approx \frac{kT}{\hbar \omega }.  \label{eq:bb}
\end{equation}
For a three-dimensional crystal (or radiation field) $g(\omega )\propto
\,\omega ^{2}$. Collecting terms, we see that the limit yielding $a^{zz}$ is
well-defined. Decoherence depends quadratically on the coupling, and
linearly on temperature and on the density of low-frequency phonons.

\subsection{Model with Dissipation}

We now generalize our model to include dissipative terms. On the other
hand, to keep the analysis tractable, we will consider the case of a single spin
coupled to a boson bath. We keep the system and bath Hamiltonians of
Eqs.~(\ref{eq:H_S}),(\ref{eq:H_B}). The new interaction Hamiltonian
is: 
\begin{eqnarray*}
{\bf H}_{I} &=&\sum_{k}{\sigma }_{z}\otimes \left( \lambda _{kz}{\bf b}
_{k}+\lambda _{kz}^{\ast }{\bf b}_{k}^{\dagger }\right) \\
&&+{\sigma }_{+}\otimes \left( \lambda _{k+}{\bf b}_{k}+\lambda
_{k+}^{\ast
} {\bf b}_{k}^{\dagger }\right) \\
&&+{\sigma }_{-}\otimes \left( \lambda _{k-}{\bf b}_{k}+\lambda _{k-}^{\ast
} {\bf b}_{k}^{\dagger }\right)
\end{eqnarray*}
(where $\lambda _{k+}=\lambda _{k-}^{\ast}$). Transforming to the IP we find: ${\bf 
\sigma }_{\alpha }(t)={\bf \sigma }_{\alpha }e^{i\omega _{0\alpha }t}$ and $ 
{\bf b}_{k}(t)={\bf b}_{k}e^{i\omega _{k}t}$ where $\alpha =z,\pm $, and $
\omega _{0z}=0$, $\omega _{0\pm }=\mp \omega _{0}.$ As above, this translates
into diagonal $p$ and $q$ [recall Eqs.~(\ref{eq:Sexp}),(\ref{eq:Bexp})]: 
\begin{equation}
p_{\alpha \beta }(t)=\delta _{\alpha \beta }e^{i\omega _{0\alpha }t}\qquad
q_{kk^{\prime }}(t)=\delta _{kk^{\prime }}e^{i\omega _{k}t},
\end{equation}
and we find for $a_{\alpha \beta }$:
\begin{eqnarray}
a_{\alpha \beta }(\tau ) &=&\frac{\tau }{\hbar ^{2}}\sum_{k}\lambda
_{k\alpha }\lambda _{k\beta }^{\ast }\langle {\bf b}_{k}^{\dagger }{\bf b}
_{k}\rangle _{B}\Gamma (\omega _{0\alpha }+\omega _{k})\Gamma (-\omega
_{0\beta }-\omega _{k})  \nonumber \\
&&+\lambda _{k\alpha ^{\prime }}^{\ast }\lambda _{k\beta ^{\prime }}\langle 
{\bf b}_{k}{\bf b}_{k}^{\dagger }\rangle _{B}\Gamma (\omega _{0\alpha
}-\omega _{k})\Gamma (-\omega _{0\beta }+\omega _{k}).  \nonumber \\
&&  \label{eq:Aab}
\end{eqnarray}
Here $\lambda _{k\alpha ^{\prime }}$ is the coupling coefficient for $\sigma
_{\alpha }^{\dagger }$, we already dropped the vanishing $\langle {\bf b}
_{k}^{\dagger }{\bf b}_{k}^{\dagger }\rangle _{B}$ and $\langle {\bf b}_{k} 
{\bf b}_{k}\rangle _{B}$ terms, and
\begin{equation}
\Gamma (\omega )=\frac{1}{\tau }\int_{0}^{\tau }e^{i\omega t}dt=e^{i\omega
\tau /2}{\rm sinc}\left( \omega \tau /2\right) .
\end{equation}
In particular, for the diagonal terms we obtain:
\begin{eqnarray}
a_{\alpha \alpha }(\tau ) &=&\frac{\tau }{\hbar ^{2}}\sum_{k}|\lambda
_{k\alpha }|^{2}\langle {\bf b}_{k}^{\dagger }{\bf b}_{k}\rangle _{B}|\Gamma
(\omega _{0\alpha }+\omega _{k})|^{2}  \nonumber \\
&&+|\lambda _{k\alpha ^{\prime }}|^{2}\langle {\bf b}_{k}{\bf b}
_{k}^{\dagger }\rangle _{B}|\Gamma (\omega _{0\alpha }-\omega _{k})|^{2}.
\end{eqnarray}
For $a_{zz}$ this yields the same result as above. For the new decoherence
parameters $a_{++}$ and $a_{--}$ we find: 
\begin{eqnarray}
a_{++}(\tau ) &=&\frac{\tau }{\hbar ^{2}}\sum_{k}|\lambda _{k+}|^{2}\left(
\langle {\bf b}_{k}^{\dagger }{\bf b}_{k}\rangle {\rm sinc}^{2}\left(
(\omega _{k}-\omega _{0})\tau /2\right) \right.  \nonumber \\
&&\left. +\langle {\bf b}_{k}{\bf b}_{k}^{\dagger }\rangle {\rm sinc}
^{2}\left( (\omega _{k}+\omega _{0})\tau /2\right) \right)  \nonumber \\
a_{--}(\tau ) &=&\frac{\tau }{\hbar ^{2}}\sum_{k}|\lambda _{k-}|^{2}\left(
\langle {\bf b}_{k}^{\dagger }{\bf b}_{k}\rangle {\rm sinc}^{2}\left(
(\omega _{k}+\omega _{0})\tau /2\right) \right.  \nonumber \\
&&\left. +\langle {\bf b}_{k}{\bf b}_{k}^{\dagger }\rangle {\rm sinc}
^{2}\left( (\omega _{k}-\omega _{0})\tau /2\right) ,\right)
\end{eqnarray}
or, using the integral form: 
\begin{eqnarray}
a_{++}(\tau ) &=&\frac{\pi }{\hbar ^{2}}\int_{0}^{\infty }d\omega \,g(\omega
)|\lambda _{+}(\omega )|^{2}\times  \nonumber \\
&&\left( \langle {\bf b}^{\dagger }{\bf b}\rangle \tilde{\delta}\left(
\omega -\omega _{0}\right) +\langle {\bf b}{\bf b}^{\dagger }\rangle \tilde{
\delta}\left( \omega +\omega _{0}\right) \right)  \nonumber \\
a_{--}(\tau ) &=&\frac{\pi }{\hbar ^{2}}\int_{0}^{\infty }d\omega \,g(\omega
)|\lambda _{-}(\omega )|^{2}\times  \nonumber \\
&&\left( \langle {\bf b}^{\dagger }{\bf b}\rangle \tilde{\delta}\left(
\omega +\omega _{0}\right) +\langle {\bf b}{\bf b}^{\dagger }\rangle \tilde{
\delta}\left( \omega -\omega _{0}\right) \right)
\end{eqnarray}
With the appearance of the system's unitary time-scale ($1/\omega _{0}$) we
have to redefine our coarse-graining procedure. We can consider two opposite
limits, where either i) the system energy, or ii) the interaction energy is
dominant. These
two limits correspond respectively to i) the system's internal unitary evolution
being fast (so we are actually coarse-graining this out as well), and ii) the
system's unitary evolution being slow.

\subsubsection{The fast system limit}

In this case $\omega _{0}\tau \gg 1$. The $\delta $ functions are centered
at $\omega _{0}$ and at $-\omega _{0}$, much further from zero than their
width. The ones at $-\omega _{0}$ thus do not contribute (off-resonance),
so, similarly to $a_{zz}$ we find:
\begin{eqnarray}
a_{++} &=&\frac{2\pi }{\hbar ^{2}}\lim_{\omega =\omega _{0}}g(\omega
)|\lambda _{+}(\omega )|^{2}\langle {\bf b}^{\dagger }{\bf b}\rangle \\
a_{--} &=&\frac{2\pi }{\hbar ^{2}}\lim_{\omega =\omega _{0}}g(\omega
)|\lambda _{-}(\omega )|^{2}\langle {\bf b}{\bf b}^{\dagger }\rangle .
\end{eqnarray}
The off-diagonal Lindblad parameters $a_{\alpha \beta }$ vanish, as they
involve the product of two $\delta $ functions that are centered further
apart than their widths. {\em The coefficient matrix is diagonal in the fast
system limit.} If we assume for further simplicity that $\lambda _{+}(\omega
)=\lambda _{-}(\omega )$, then the diagonal parameters $a_{++}$ and $a_{--}$
only differ in the bath expectation values at $\omega _{0}$. It then follows
that
\begin{equation}
a_{--} = a_{++}e^{\beta\hbar\omega_{0}}.
\end{equation}

Let us now consider briefly the resulting interaction-picture master
equation. Using the notation $a=a_{zz},b=a_{++},c=a_{--}$ for simplicity, we
obtain:
\begin{equation}
\frac{d\rho }{dt}=\left( 
\begin{array}{cc}
c\rho _{11}-b\rho _{00} & -(2a+\frac{b+c}{2})\rho _{01} \\ 
-(2a+\frac{b+c}{2})\rho _{10} & b\rho _{00}-c\rho _{11}
\end{array}
\right)
\end{equation}
The off-diagonal elements decay exponentially, with a rate $\tau _{{\rm dec}
}^{-1}=(2a+\frac{b+c}{2})$. The diagonal elements approach the thermal
equilibrium values $\rho _{00}^{{\rm ther}}$ and $\rho _{11}^{{\rm ther}}$,
where
\begin{equation}
\frac{\rho _{00}^{{\rm ther}}}{\rho _{11}^{{\rm ther}}}=\frac{c}{b}=\frac{
a_{--}}{a_{++}}=e^{\beta \hbar \omega _{0}}.
\end{equation}
The exponential rate of convergence of the diagonal elements, i.e., the
dissipation rate, is $\tau _{{\rm diss}}^{-1}=a_{--}+a_{++}$. Within the
framework of our model, both rates depend linearly on temperature and
quadratically on the corresponding coupling strengths. The important
difference between them is the presence of $a_{zz}$ in the dephasing rate.
The parameters $a_{++}$ and $a_{--}$ depend on the bath's density of states
at $\omega _{0}$. Dissipation therefore can be quite slow in a number of
important cases, for example when there is a gap in the phonon spectrum, or
when $\omega _{0}$ is actually greater than the cutoff frequency. In these
cases only much weaker multi-phonon processes cause dissipation.

The parameter $a_{zz}$, on the other hand, depends on the density of
low-frequency phonons. This can be small only in very special circumstances
(e.g., superfluidity, or a discrete phonon density of states as would be
found in a quantum dot \cite{Takagahara:96}) and its vanishing indeed
usually causes macroscopic quantum-effects. In typical situations the rate
of dephasing will be greater than the rate of dissipation.

The important general conclusion is the following: If our coarse-graining
includes the (fast) system as well, then the density matrix rapidly
decoheres into the system's energy eigen-basis \cite{Paz:99}. Then,
(typically slower) it converges into the {\em thermalized} density matrix
(which is of course also diagonal in the system's energy eigen-basis). See 
\cite{Palma:96} for a more detailed discussion of these different regimes.

\subsubsection{The slow system limit}

In this case $\omega _{0}\tau \ll 1$. We consider only the zeroth
approximation, i.e., set $\omega _{0}=0$. Using Eq.~(\ref{eq:Aab}) we
obtain: 
\begin{eqnarray*}
a_{\alpha \beta }(\tau ) &=&\frac{\tau }{\hbar ^{2}}\sum_{k}(\lambda
_{k\alpha }\lambda _{k\beta }^{\ast }\langle {\bf b}_{k}^{\dagger }{\bf b}
_{k}\rangle \\
&&+\lambda _{k\alpha }^{\ast }\lambda _{k\beta }\langle {\bf b}_{k}{\bf b}
_{k}^{\dagger }\rangle )|\Gamma (\omega _{k})|^{2},
\end{eqnarray*}
or, integrating out the $\delta $ function again, using Eq.~(\ref{eq:bb})
for the present low-frequency limit, and assuming real $\lambda _{k\alpha }$
for simplicity: 
\begin{eqnarray}
a_{\alpha \beta } &=&\frac{2\pi }{\hbar ^{2}}kT\lim_{\omega =0}\frac{
g(\omega )}{\hbar \omega }\lambda _{\alpha }(\omega )\lambda _{\beta
}(\omega )=a_{\alpha }a_{\beta }  \nonumber \\
a_{\alpha } &\equiv &\left( \frac{2\pi }{\hbar ^{2}}kT\lim_{\omega =0}\frac{
g(\omega )}{\hbar \omega }\lambda _{\alpha }(\omega )^{2}\right) ^{1/2}.
\end{eqnarray}
Unlike in the fast-system case, the off-diagonal elements of the coefficient
matrix do not vanish. Instead, {\em in the slow-system limit} $a_{\alpha
\beta }$ is a {\em projection}, i.e., $a_{\alpha \beta }$ is an outer
product of the vector of components $\{a_{\alpha }\}$ with itself. This
allows us to write the SME using just one Lindblad operator: 
\begin{equation}
{\bf G}\equiv \sum_{\alpha }a_{\alpha }\sigma _{\alpha }
\end{equation}
This course-grained interaction operator is just a linear combination of the
system operators as they appear in the interaction Hamiltonian, but with the
dependence on the bath degrees of freedom already averaged out. Using ${\bf 
G }$, the SME becomes:
\begin{equation}
\frac{\partial \rho (t)}{\partial t}=\frac{1}{2}([{\bf G},\rho (t){\bf G}
^{\dagger }]+[{\bf G}\rho (t),{\bf G}^{\dagger }])
\end{equation}
Diagonalizing ${\bf G}$ and transforming $\rho $ into ${\bf G}$'s eigenbasis
then leads to uncoupled equations for the components of the transformed $
\rho $. Therefore, in the slow system limit, the density matrix becomes
diagonal in the eigenbasis of the course-grained interaction Hamiltonian
(i.e., ${\bf G}$), and for the rate of this decoherence we find:
\begin{equation}
\tau _{{\rm dec}}^{-1}=\frac{2\pi }{\hbar ^{2}}kT\lim_{\omega =0}\left( 
\frac{g(\omega )}{\omega }\left( 2\lambda _{z}^{2}(\omega )+\lambda
_{+}^{2}(\omega )+\lambda _{-}^{2}(\omega )\right) \right) .
\end{equation}

\subsection{Comparison of the Markovian Result to Exact Solution of
Spin-Boson Model for Pure Dephasing}

The spin-boson model is exactly solvable in the pure dephasing limit, and we
present the detailed solution in Appendix \ref{appA}. The result for an
initial thermal bath is that the time-dependence of the off-diagonal terms
is proportional to $e^{-\Gamma (T,t)}$, with 
\begin{equation}
\Gamma =\frac{2t^{2}}{\hbar ^{2}}\sum_{k}|\lambda _{k}|^{2}{\rm sinc}
^{2}\left( \omega _{k}t/2\right) \coth \frac{\hbar \omega _{k}}{2k_{B}T}.
\label{eq:Gamma(t)}
\end{equation}
This exact result holds for arbitrary times $t$ and for both finite and
infinite baths.

On the other hand, recall that the SME result for the single qubit case was
[Eq.~(\ref{eq:aijSB})]: 
\begin{equation}
a_{zz}=\frac{\tau }{\hbar ^{2}}\sum_{k}|\lambda _{k}|^{2}{\rm sinc}
^{2}\left( \omega _{k}\tau /2\right) \coth \frac{\hbar \omega _{k}}{2k_{B}T}
\end{equation}
This is the dephasing rate for a single qubit satisfying the Lindblad master
equation 
\begin{equation}
\frac{d\rho }{dt}=\frac{1}{2}a_{zz}\left( \left[ \sigma _{z}\rho ,\sigma
_{z} \right] +\left[ \sigma _{z},\rho \sigma _{z}\right] \right) ,
\end{equation}
whence the off-diagonal $\rho _{01}\propto \exp (-2a_{zz}t)$.

How do these results relate to one another? We have in the Markovian case: 
\begin{equation}
\rho _{01}^{{\rm SME}}\propto \exp \left( -t\frac{2\pi }{\hbar ^{2}}
\sum_{k}|\lambda _{k}|^{2}\tilde{\delta}(\tau ,\omega _{k})\coth \frac{\hbar
\omega _{k}}{2k_{B}T}\right)  \label{eq:SMEfinal}
\end{equation}
whereas in the exact case: 
\begin{equation}
\rho _{01}^{{\rm exact}}\propto \exp \left( -t\frac{2\pi }{\hbar ^{2}}
\sum_{k}|\lambda _{k}|^{2}\tilde{\delta}(t,\omega _{k})\coth \frac{\hbar
\omega _{k}}{2k_{B}T}\right).  \label{eq:exactfinal}
\end{equation}
While superficially the similarity between these results is striking, there
is nevertheless a crucial difference: The exact solution has {\em recurrences},
since its time-dependence is periodic (for a {\em finite} bath), whereas the SME
result is a purely exponential decay. Thus they describe very different
behaviors. Indeed, for small $t$ ($\omega _{k}t\ll 1$) the exact result
decays as $\exp (-t^{2})$ (quantum Zeno effect \cite{Misra:77,Itano:90}), while the Markovian result
always decays as $\exp (-t)$. This is of course not a surprise: the
Markovian result cannot describe the dynamics for times shorter than the
coarse-graining time-scale, $\tau$.

Let us now turn to see the limit in which the two solutions do agree. To
prevent recurrences in the exact solution we once again replace the sum over
modes by an integral, to obtain: 
\begin{eqnarray*}
\rho _{01}^{{\rm SME}} &\propto &\exp \left( -t\frac{2\pi }{\hbar ^{2}}\int
d\omega g(\omega )|\lambda (\omega )|^{2}\tilde{\delta}(\tau ,\omega )\coth 
\frac{\hbar \omega }{2k_{B}T}\right) \\
\rho _{01}^{{\rm exact}} &\propto &\exp \left( -t\frac{2\pi }{\hbar ^{2}}
\int d\omega g(\omega )|\lambda (\omega )|^{2}\tilde{\delta}(t,\omega )\coth 
\frac{\hbar \omega }{2k_{B}T}\right) .
\end{eqnarray*}
The only difference is the appearance of $\tau $ and $t$ in the widths of
the $\tilde{\delta}$ functions. Now, our coarse-graining procedure was
defined such that $\tau \gg 1/\omega _{c}$ [recall the discussion
surrounding Eq.~(\ref{eq:deltafunc})], and in this limit $\tilde{\delta}
(\tau ,\omega )=\delta (\omega )$. For times $t>\tau $, $\tilde{\delta}
(t,\omega )=\delta (\omega )$ also holds, so we can summarize the condition
for the exact and Markovian solutions to agree as: 
\begin{equation}
t>\tau \gg 1/\omega _{c}.  \label{eq:conds}
\end{equation}
To illustrate this let us consider the Debye model as a simple example.
Then: 
\[
g(\omega )\propto \left\{ 
\begin{array}{c}
\omega ^{2}\text{ for }\omega <\omega _{c} \\ 
0\text{ for }\omega \geq \omega _{c}
\end{array}
\right. . 
\]
As before, let the coupling coefficient $\lambda $ depend on $\omega $ only
due to amplitude-coupling: $|\lambda (\omega )|^{2}\propto \omega ^{-1}$. In
the high-temperature limit $\coth (\frac{\hbar \omega }{2k_{B}T})\propto
\omega ^{-1}$, so that in all we have
\begin{eqnarray}
\rho _{01}^{{\rm SME}} &\propto &\exp \left( -Ct\tau \int_{0}^{\omega
_{c}}d\omega {\rm sinc}^{2}\left( \omega \tau /2\right) \right)
\label{eq:SME-Debye} \\
\rho _{01}^{{\rm exact}} &\propto &\exp \left( -Ct^{2}\int_{0}^{\omega
_{c}}d\omega {\rm sinc}^{2}\left( \omega t/2\right) \right) ,
\label{eq:exact-Debye}
\end{eqnarray}
where $C$ is the temperature-dependent coupling-strength, with dimensions of
frequency. Figure 1 shows the argument of the exponential, $\Gamma (t)$, for
the exact solution and for the SME results, corresponding to different
values of the course-graining time-scale, $\tau $. The curves corresponding
to the SME solutions of course are just straight lines, as they all describe
simple exponential decays. It is clear that the SME solutions cannot account
for the initial transition period, but for sufficiently large $\tau $
(in units of the bath cut-off time $1/\omega_c$) the
SME result approximates the exact solution very well at large times, in
accordance with Eq.~(\ref{eq:conds}).

Let us summarize: The Markovian approximation we introduced gives reliable
results for times greater than the course-graining time-scale, which in turn
must be greater than the bath cut-off time. It does not account for the
initial (transitional) time evolution, and it should be applied in cases of
an infinite bath with continuous spectrum.

\subsection{The Lamb Shift}

Finally, in the exact solution for {\em multiple} qubits there is also a
non-vanishing Lamb shift, which arises as a consequence of the Hamiltonian
not commuting with itself at different times \cite{Duan:97}. The Lamb shift
does vanish for a single qubit in the exact solution of the pure dephasing
spin-boson model \thinspace (see Appendix \ref{appA} and \cite{Duan:97}).
The Lamb shift also vanished in our Markovian calculation. This discrepancy
is not only due to the fact that we considered a single qubit:\ the more
fundamental reason is that we only carried out our Markovian calculations to
first order in perturbation theory, where time-ordering did not play a role.
However, when we consider the multiple-qubit case in {\em second}-order
perturbation theory (recall Section \ref{2ndorder}) there {\em is} a
Lamb-shift. This arises because of terms like $\sigma _{z}^{i}\sigma
_{z}^{j}b_{k}^{\dagger }b_{k}$. Physically, this is a phonon-induced,
indirect, exchange-interaction between the two spins. It is quadratic in $
\lambda $, linear in temperature, and acts to pull the spin-energies towards
an average value.

\section{Conclusions}
\label{conclusions}

A central task of modern condensed phase chemistry and physics is the
quantitative description of open quantum systems. These are systems that are
coupled to an external {\em uncontrollable} environment (bath), a coupling
which generally leads to decoherence. In this paper we provided such a
quantitative description, by deriving a practical way to calculate the
coefficients in the quantum Markovian semigroup master equation (commonly
known as the Lindblad equation). Our starting point was the exact Kraus
operator sum representation, which presents the evolution of an open quantum
system as a general, completely positive, linear map. By coarse-graining
this evolution over a time-scale typical of the bath (the inverse of the
bath density-of-states frequency-cutoff), we showed how the Lindblad
equation can be derived, and how its coefficients can be systematically calculated 
using perturbation theory in the system-bath coupling strength. This
resolves an important shortcoming in the theory of open quantum systems: so
far no practical general method was known which takes as input an
interaction Hamiltonian, and then produces the Lindblad equation {\it together}
with all its
coefficients. The complexity of our method is determined by the difficulty
of calculating certain time-ordered integrals, which of course increases
with higher orders of perturbation theory. In principle, this is equivalent
to the calculation of standard Feynman diagrams, and thus the arsenal of
techniques known in many-body physics could be employed here as well. To test
the validity of our theory, we compared it here to an exactly solvable model,
namely, the
spin-boson Hamiltonian with pure phase-damping. For times longer than the
coarse-graining time, the agreement was found to be excellent already at the
level of first order perturbation theory. 

\section{Acknowledgements}

It is a pleasure to acknowledge very insightful discussions with Dave Bacon.
This work was supported in part by the National Security Agency (NSA)
and Advanced Research and Development Activity (ARDA) under Army Research
Office (ARO) contract number DAAG55-98-1-0371.

\appendix

\section{Analytical Solution of the Spin Boson Model for Pure Dephasing}
\label{appA}

We present here the analytical solution of the spin-boson model for pure
dephasing. The derivation is based on \cite{Duan:97}.

Starting from the interaction picture Hamiltonian:
\begin{equation}
{\bf H}_{I}(t)=\sum_{i,k}{\sigma }_{z}^{i}\otimes \left[ \lambda
_{k}^{i}e^{-i\omega _{k}t}{\bf a}_{k}+\left( \lambda _{k}^{i}\right) ^{\ast
}e^{i\omega _{k}t}{\bf a}_{k}^{\dagger }\right]
\end{equation}
we want to find the system density matrix 
\begin{equation}
\rho _{I}(t)={\rm Tr}_{B}\left[ \rho _{{\rm tot,}I}(t)\right] ={\rm Tr}_{B} 
\left[ {\bf U}(t)\rho (0)\otimes \rho _{B}(0){\bf U}^{\dagger }(t)\right]
\end{equation}
where 
\[
{\bf U}(t)={\rm T}\exp \left[ -\frac{i}{\hbar }\int_{0}^{t}{\bf H}_{I}(\tau
)d\tau \right] . 
\]

\subsection{Calculation of the Evolution Operator}

Note that ${\bf H}_{I}(t)$ does not commute with itself at different times,
which is why we need the time-ordered product: 
\[
\left[ {\bf H}_{I}(t),{\bf H}_{I}(t^{\prime })\right] =\sum_{i,i^{\prime
};k}2i{\sigma }_{z}^{i}{\sigma }_{z}^{i^{\prime }}\otimes {\rm Re}\left[
\lambda _{k}^{i}\left( \lambda _{k}^{i^{\prime }}\right) ^{\ast }\right]
\sin \omega _{k}\left( t-t^{\prime }\right) {\bf 1,}
\]
where we used the boson commutation relations $\left[ {\bf a}_{k},{\bf a}
_{l}^{\dagger }\right] ={\bf 1}\delta _{kl}$, $\left[ {\bf a}_{k},{\bf a}_{l}
\right] =0$. Note that further, 
\begin{equation}
\left[ \left[ {\bf H}_{I}(t),{\bf H}_{I}(t^{\prime })\right] ,{\bf H}
_{I}(t^{\prime \prime })\right] =0.
\end{equation}
This means that we can use the Baker-Hausdorf formula $\exp (A+B)=\exp
(-[A,B]/2)\exp (A)\exp (B)$ (valid if $[[A,B],A]=[[A,B],B]=0$) to calculate $
{\bf U}(t)$. To do so note the generalization 
\begin{equation}
\exp \left( \sum_{n}A_{n}\right) =\left( \prod_{n<n^{\prime }}\exp \left( -
\frac{1}{2}[A_{n},A_{n^{\prime }}]\right) \right) \left( \prod_{n}\exp
(A_{n})\right) 
\end{equation}
which is valid if every second-order commutator vanishes. To apply this
result for our case let us formally discretize the integrals and denote $
{\cal H}_{n}\equiv -\frac{i}{\hbar }{\bf H}_{I}(n\Delta t)$. Then: 
\begin{eqnarray*}
{\bf U}(t) &=&{\rm T}\exp \left[ -\frac{i}{\hbar }\int_{0}^{t}{\bf H}
_{I}(\tau )d\tau \right] ={\rm T}\lim_{\Delta t\rightarrow 0}\exp \left[
\sum_{n=0}^{N}{\cal H}_{n}\Delta t\right]  \\
&=&\lim_{\Delta t\rightarrow 0}\prod_{n<n^{\prime }}\exp \left( -\frac{1}{2}[
{\cal H}_{n},{\cal H}_{n^{\prime }}]\left( \Delta t\right) ^{2}\right)
\prod_{n}\exp ({\cal H}_{n}\Delta t) \\
&=&\lim_{\Delta t\rightarrow 0}\prod_{n<n^{\prime }}\left( 1-\frac{1}{2}[
{\cal H}_{n},{\cal H}_{n^{\prime }}]\left( \Delta t\right) ^{2}\right)
\prod_{n}\left( 1-{\cal H}_{n}\Delta t\right)  \\
&=&\lim_{\Delta t\rightarrow 0}\left[ 1-\frac{1}{2}\sum_{n<n^{\prime }}[
{\cal H}_{n},{\cal H}_{n^{\prime }}]\left( \Delta t\right) ^{2}\right] \left[
1-\sum_{n}{\cal H}_{n}\Delta t\right]  \\
&=&\lim_{\Delta t\rightarrow 0}\exp \left( -\frac{1}{2}\sum_{n<n^{\prime }}[
{\cal H}_{n},{\cal H}_{n^{\prime }}]\left( \Delta t\right) ^{2}\right) \exp
( \sum_{n}{\cal H}_{n}\Delta t )  \\
&=&\exp \left[ \left( -\frac{i}{\hbar }\right)
^{2}\int_{0}^{t}dt_{1}\int_{0}^{t_{1}}dt_{2}\left[ {\bf H}_{I}(t_{2}),{\bf H}
_{I}(t_{1})\right] \right] \times  \\
&&\exp \left[ -\frac{i}{\hbar }\int_{0}^{t}{\bf H}_{I}(\tau )d\tau \right] .
\end{eqnarray*}
In the second line we explicitly invoked time-ordering by using the
Campbell-Hausdorf formula to deal with the non-commuting problem; in the
subsequent lines we used the re-exponentiation trick. In the final line
there is no need for explicit time-ordering left, i.e., the integrals can be
calculated as such. We find: 
\begin{equation}
-\frac{i}{\hbar }\int_{0}^{t}{\bf H}_{I}(\tau )d\tau ={\sigma }
_{z}^{i}\otimes \sum_{i,k}\left( \alpha _{k}^{i}(t){\bf a}_{k}^{\dagger
}-\alpha _{k}^{i}(t)^{\ast }{\bf a}_{k}\right) 
\end{equation}
where
\begin{equation}
\alpha _{k}^{i}(t)=\frac{\left( \lambda _{k}^{i}\right) ^{\ast }(e^{i\omega
_{k}t}-1)}{\hbar \omega _{k}}.
\end{equation}
Further: 
\begin{equation}
\int_{0}^{t}dt_{1}\int_{0}^{t_{1}}dt_{2}\sin \omega _{k}\left(
t_{2}-t_{1}\right) =\frac{\sin \omega _{k}t-\omega _{k}t}{\omega _{k}^{2}},
\end{equation}
and 
\begin{equation}
\int_{0}^{t}dt_{1}\int_{0}^{t_{1}}dt_{2}\cos \omega _{k}\left(
t_{2}-t_{1}\right) =\frac{1-\cos \omega _{k}t}{\omega _{k}^{2}},
\end{equation}
so that 
\begin{eqnarray}
f(t)\equiv - &&i\left( -\frac{i}{\hbar }\right)
^{2}\int_{0}^{t}dt_{1}\int_{0}^{t_{1}}dt_{2}\left[ {\bf H}_{I}(t_{2}),{\bf H}
_{I}(t_{1})\right]  \nonumber \\
&=&\sum_{i,i^{\prime };k}2{\sigma }_{z}^{i}{\sigma }_{z}^{i^{\prime
}}\otimes {\rm Re}\left[ \lambda _{k}^{i}\left( \lambda _{k}^{i^{\prime
}}\right) ^{\ast }\right] \frac{\omega _{k}t-\sin \omega _{k}t}{\left( \hbar
\omega _{k}\right) ^{2}}{\bf 1}  \nonumber \\
&=&\sum_{k}\frac{\omega _{k}t-\sin \omega _{k}t}{\left( \hbar \omega
_{k}\right) ^{2}}\left| \sum_{i}\lambda _{k}^{i}{\sigma }_{z}^{i}\right|
^{2}\otimes {\bf 1}.
\end{eqnarray}

Note that $f$ is an operator acting just on the system, and is a simple
phase for the case of a single qubit. Since the ${\bf a}_k$ operators
commute for different modes we have as our final simplified result for the
evolution operator: 
\begin{equation}
{\bf U}(t)=e^{if(t)}\prod_{i,k}\exp \left[ {\sigma }_{z}^{i}\otimes \left(
\alpha _{k}^{i}(t){\bf a}_{k}-\alpha _{k}^{i}(t)^{\ast }{\bf a}_{k}^{\dagger
}\right) \right] .
\end{equation}

\subsection{Calculation of the Density Matrix}

Now recall the definition of the coherent states. These are eigenstates of
the annihilation operator: 
\begin{equation}
{\bf a}|\alpha \rangle =\alpha |\alpha \rangle .
\end{equation}
They are minimum-uncertainty states in a harmonic potential, etc. As is well
known, 
\begin{equation}
|\alpha \rangle =e^{-|\alpha |^{2}/2}\sum_{n=0}^{\infty }\frac{\alpha ^{n}}{
\sqrt{n!}}|n\rangle
\end{equation}
where $|n\rangle $ are number (Fock) states. The completeness relation for
the coherent states is: 
\begin{equation}
\frac{1}{\pi }\int d^{2}\alpha \,|\alpha \rangle \langle \alpha |=1
\end{equation}
where the integration is over the entire complex plane. They are useful in
our context since they are created by the displacement operator 
\begin{equation}
D\left( \alpha \right) \equiv \exp \left( \alpha {\bf a}^{\dagger }-\alpha
^{\ast }{\bf a}\right) =D(-\alpha )^{\dagger }
\end{equation}
acting on the vacuum state: 
\begin{equation}
D\left( \alpha \right) |{\bf 0}\rangle =|\alpha \rangle ,
\end{equation}
which is clearly related to ${\bf U}(t)$. We will need the result: 
\begin{equation}
D\left( \alpha \right) D\left( \beta \right) =\exp \frac{\alpha \beta ^{\ast
}-\alpha ^{\ast }\beta }{2}D(\alpha +\beta ),
\end{equation}
which is easily derived from $D\left( \alpha \right) =\exp \left( \alpha 
{\bf a}^{\dagger }-\alpha ^{\ast }{\bf a}\right) $, $[{\bf a,a}^{\dagger
}]=1 $, and the Baker-Hausdorf formula $\exp (A+B)=\exp (-[A,B]/2)\exp
(A)\exp (B) $ (again, valid if $[[A,B],A]=[[A,B],B]=0$).

Now let $R_{ik}(t)\equiv \alpha _{k}^{i}(t){\bf a}_{k}^{\dagger }-\alpha
_{k}^{i}(t)^{\ast }{\bf a}_{k}$ and consider $\exp \left[ {\sigma }
_{z}^{i}\otimes R_{ik}(t)\right] $: 
\begin{eqnarray}
\exp \left[ {\sigma }_{z}\otimes R\right]  &=&I_{S}\otimes
\sum_{n=0}^{\infty }\frac{R^{2n}}{(2n)!}+{\sigma }_{z}\otimes
\sum_{n=0}^{\infty }\frac{R^{2n+1}}{(2n+!)!}  \nonumber \\
&=&I_{S}\otimes \cosh R+{\sigma }_{z}\otimes \sinh R  \nonumber \\
&=&I_{S}\otimes \frac{1}{2}[D\left( \alpha \right) +D\left( -\alpha \right) ]
\nonumber \\
&&+{\sigma }_{z}\otimes \frac{1}{2}[D\left( \alpha \right) -D\left( -\alpha
\right) ]  \nonumber \\
&=&|0\rangle \langle 0|\otimes D\left( \alpha \right) +|1\rangle \langle
1|\otimes D\left( -\alpha \right) .
\end{eqnarray}
This is an important result since it shows that depending on whether the
field is coupled to the qubit $|0\rangle $ or $|1\rangle $ state, the field
acquires a different displacement. This is the source of the dephasing the
qubits undergo, since when acting on a superposition state of a qubit, the
qubit and field become entangled: 
\begin{eqnarray}
&& \exp \left[ {\sigma }_{z}\otimes R\right] (a|0\rangle +b|1\rangle )|\beta
\rangle  =a|0\rangle \otimes D\left( \alpha \right) |\beta \rangle  
\nonumber \\
&&+b|1\rangle \otimes D\left( -\alpha \right) |\beta \rangle   \nonumber \\
&=&e^{(\alpha \beta ^{\ast }-\alpha ^{\ast }\beta )/2}a|0\rangle \otimes
|\alpha +\beta \rangle   \nonumber \\
&&+e^{-(\alpha \beta ^{\ast }-\alpha ^{\ast }\beta )/2}b|1\rangle \otimes
|\beta -\alpha \rangle .\nonumber 
\end{eqnarray}
The evolution operator can be written as: 
\begin{equation}
{\bf U}(t)=e^{if(t)}\prod_{i,k}\left[ |0\rangle _{i}\langle 0|\otimes
D\left( \alpha _{k}^{i}\right) +|1\rangle _{i}\langle 1|\otimes D\left(
-\alpha _{k}^{i}\right) \right] .
\end{equation}
Now assume that the boson bath is in thermal equilibrium: 
\begin{eqnarray}
\rho _{B} &=&\frac{1}{Z}e^{-\beta {\bf H}_{B}}  \nonumber \\
&=&\left[ \prod_{k}\frac{e^{-\beta \hbar \omega _{k}/2}}{1-e^{-\beta \hbar
\omega _{k}}}\right] ^{-1}\exp \left( -\beta \sum_{k}\hbar \omega _{k}\left( 
{\bf N}_{k}+\frac{1}{2}\right) \right)   \nonumber \\
&=&\prod_{k}\frac{1}{\langle {\bf N}_{k}\rangle }\exp \left( -\beta \hbar
\omega _{k}{\bf N}_{k}\right) ,
\end{eqnarray}
where the mean boson occupation number is: 
\begin{equation}
\langle {\bf N}_{k}\rangle =\frac{1}{e^{\beta \hbar \omega _{k}}-1}.
\end{equation}
As shown in \cite{Gardiner:book}, p.122-3, this can be transformed into the
coherent-state representation, with the result: 
\begin{equation}
\rho _{B}=\prod_{k}\rho _{B,k}
\end{equation}
where 
\begin{equation}
\rho _{B,k}=\frac{1}{\pi \langle {\bf N}_{k}\rangle }\int d^{2}\alpha
_{k}\,\exp \left( -\frac{|\alpha _{k}|^{2}}{\langle {\bf N}_{k}\rangle }
\right) |\alpha _{k}\rangle \langle \alpha _{k}|.
\end{equation}

Now consider the system density matrix. Let $\rho _{x_{i},y_{i}}=|x\rangle
_{i}\langle y|$ where $x,y=\{0,1\}$. Since we are dealing with qubits the
system density matrix is a sum of all possible tensor products of single
qubit pure states, i.e., of terms of the form $\rho _{\{x_{i},y_{i}\}}\equiv
\rho _{x_{1},y_{1}}\otimes \cdots \otimes \rho _{x_{N},y_{N}}$. Thus it\ can
be expanded as 
\begin{equation}
\rho (0)=\sum_{\{x_{i},y_{i}\}}c_{\{x_{i},y_{i}\}}\rho _{\{x_{i},y_{i}\}}.
\end{equation}
Recall that we set out to evaluate $\rho (t)={\rm Tr}_{B}\left[ {\bf U}
(t)\rho (0)\otimes \rho _{B}(0){\bf U}^{\dagger }(t)\right] $. For
simplicity let us now consider the case of a single qubit. It suffices to
calculate the evolution of each of the four pure states $|x\rangle \langle
y| $ separately. Thus
\begin{eqnarray*}
\rho _{x,y}(t) &=&{\rm Tr}_{B}\left[ {\bf U}(t)|x\rangle \langle y|\otimes
\rho _{B}(0){\bf U}^{\dagger }(t)\right]  \\
&=&{\rm Tr}_{B}\left[ e^{if(t)}\prod_{k}\left[ |0\rangle \langle 0|\otimes
D\left( \alpha _{k}\right) +|1\rangle \langle 1|\otimes D\left( -\alpha
_{k}\right) \right] \right.  \\
&&\left. |x\rangle \langle y|\otimes \prod_{m}\rho _{B,m}\right.  \\
&&\left. \prod_{l}\left[ |0\rangle \langle 0|\otimes D^{\dagger }\left(
\alpha _{l}\right) +|1\rangle \langle 1|\otimes D^{\dagger }\left( -\alpha
_{l}\right) \right] e^{-if^{\dagger }(t)}\right] .
\end{eqnarray*}
The terms in the three products match one-to-one for equal indices, so we
can write everything as a product over a single index $k$. Using ${\rm Tr}
(A\otimes B)={\rm Tr}A\times {\rm Tr}B$ to rearrange the order of the trace
and the products, and $D^{\dagger }\left( -\alpha \right) =D\left( \alpha
\right) $, we have:

\begin{eqnarray*}
\rho _{x,y}(t) &=&\delta _{x,0}\delta _{y,0}e^{if(t)}|0\rangle \langle
0|e^{-if^{\dagger }(t)} \\
&&\otimes \prod_{k}{\rm Tr}_{B}\left[ D\left( \alpha _{k}\right) \rho
_{B,k}D\left( -\alpha _{k}\right) \right]  \\
&+&\delta _{x,0}\delta _{y,1}e^{if(t)}|0\rangle \langle 1|e^{-if^{\dagger
}(t)} \\
&&\otimes \prod_{k}{\rm Tr}_{B}\left[ D\left( \alpha _{k}\right) \rho
_{B,k}D\left( \alpha _{k}\right) \right]  \\
&+&\delta _{x,1}\delta _{y,0}e^{if(t)}|1\rangle \langle 0|e^{-if^{\dagger
}(t)} \\
&&\otimes \prod_{k}{\rm Tr}_{B}\left[ D\left( -\alpha _{k}\right) \rho
_{B,k}D\left( -\alpha _{k}\right) \right]  \\
&+&\delta _{x,1}\delta _{y,1}e^{if(t)}|1\rangle \langle 1|e^{-if^{\dagger
}(t)} \\
&&\otimes \prod_{k}{\rm Tr}_{B}\left[ D\left( -\alpha _{k}\right) \rho
_{B,k}D\left( \alpha _{k}\right) \right] .
\end{eqnarray*}
Consider the ${\rm Tr}_{B}$ terms:\ for $|0\rangle \langle 0|$ and $
|1\rangle \langle 1|$ by cycling in the trace the displacement operators
cancel and ${\rm Tr}_{B}\left[ \rho _{B,k}\right] =1$. Thus, as expected the
\ diagonal terms do not change [apart from the Lamb shift due to $f(t)$]. As
for the off-diagonal terms (evaluating the trace in any complete basis): 
\begin{eqnarray*}
&&{\rm Tr}_{B}\left[ D\left( \pm 2\alpha _{k}\right) \rho
_{B,k}\right]  \nonumber \\
&=& \frac{1}{\pi \langle {\bf N}_{k}\rangle }\int d^{2}\beta _{k}\,\exp \left( -
\frac{|\beta _{k}|^{2}}{\langle {\bf N}_{k}\rangle }\right)
\sum_{n}\langle n|D\left( \pm 2\alpha _{k}\right) |\beta _{k}\rangle
\langle \beta _{k}|n\rangle  \\
&=&\frac{1}{\pi \langle {\bf N}_{k}\rangle }\int d^{2}\beta _{k}\,\exp
\left( -\frac{|\beta _{k}|^{2}}{\langle {\bf N}_{k}\rangle }\right) 
\langle \beta _{k}|D\left( \pm 2\alpha _{k}\right) |\beta _{k}\rangle .
\end{eqnarray*}
Now: 
\begin{eqnarray*}
&&\langle \beta |D\left( \pm 2\alpha \right) |\beta \rangle  \nonumber
\\
&=&\exp \left[
\pm \left( \alpha \beta ^{\ast }-\alpha ^{\ast }\beta \right) \right]
\langle \beta |\pm \alpha +\beta \rangle  \\
&=&\exp \left[ \pm \left( \alpha \beta ^{\ast }-\alpha ^{\ast }\beta \right) 
\right] \times  \\
&&\exp \left[ \beta ^{\ast }\left( \pm 2\alpha +\beta \right) -\frac{1}{2}
\left( |\beta |^{2}+|\pm 2\alpha +\beta |^{2}\right) \right]  \\
&=&\exp \left( -2|\alpha |^{2}\pm 2\left( \alpha \beta ^{\ast }-\alpha
^{\ast }\beta \right) \right) .
\end{eqnarray*}
Thus: 
\begin{eqnarray*}
&&{\rm Tr}_{B}\left[ D \left( \pm 2\alpha _{k}\right) \rho _{B,k}\right] 
=\exp \left( -2|\alpha _{k}|^{2}\right) \frac{1}{\pi \langle {\bf N}
_{k}\rangle }\times  \\
&&\int d^{2}\beta _{k}\,\exp \left( -\frac{|\beta _{k}|^{2}}{\langle {\bf N}
_{k}\rangle }\pm 2\left( \alpha _{k}\beta _{k}^{\ast }-\alpha _{k}^{\ast
}\beta _{k}\right) \right)  \\
&=& \frac{\exp \left( -2|\alpha _{k}|^{2}\right)}{\pi \langle {\bf N}
_{k}\rangle }\left[ \pi \langle {\bf N}_{k}\rangle \exp \left( -4|\alpha
_{k}|^{2}\langle {\bf N}_{k}\rangle \right) \right]  \\
&=&\exp \left[ -4|\alpha _{k}|^{2}\left( \langle {\bf N}_{k}\rangle +\frac{1
}{2}\right) \right]  \\
&=&\exp \left[ -4\left| \frac{\lambda _{k}^{\ast }(e^{i\omega _{k}t}-1)}{
\hbar \omega _{k}}\right| ^{2}\left( \frac{1}{e^{\beta \hbar \omega _{k}}-1}+
\frac{1}{2}\right) \right]  \\
&=&\exp \left[ -4|\lambda _{k}|^{2}\frac{1-\cos \omega _{k}t}{\left( \hbar
\omega _{k}\right) ^{2}}\coth \beta \hbar \omega _{k}/2\right] .
\end{eqnarray*}
Thus decay of the off-diagonal terms goes as $e^{-\Gamma (T,t)}$, with 
\begin{equation}
\Gamma =4\sum_{k}|\lambda _{k}|^{2}\frac{1-\cos \omega _{k}t}{\left( \hbar
\omega _{k}\right) ^{2}}\coth \frac{\hbar \omega _{k}}{2k_{B}T},
\end{equation}
which is equivalent to the result that appeared in Eq.(\ref{eq:Gamma(t)})
above.

\end{multicols}

\begin{figure}[!htb]
\hspace{0.2\textwidth}
\psfig{figure=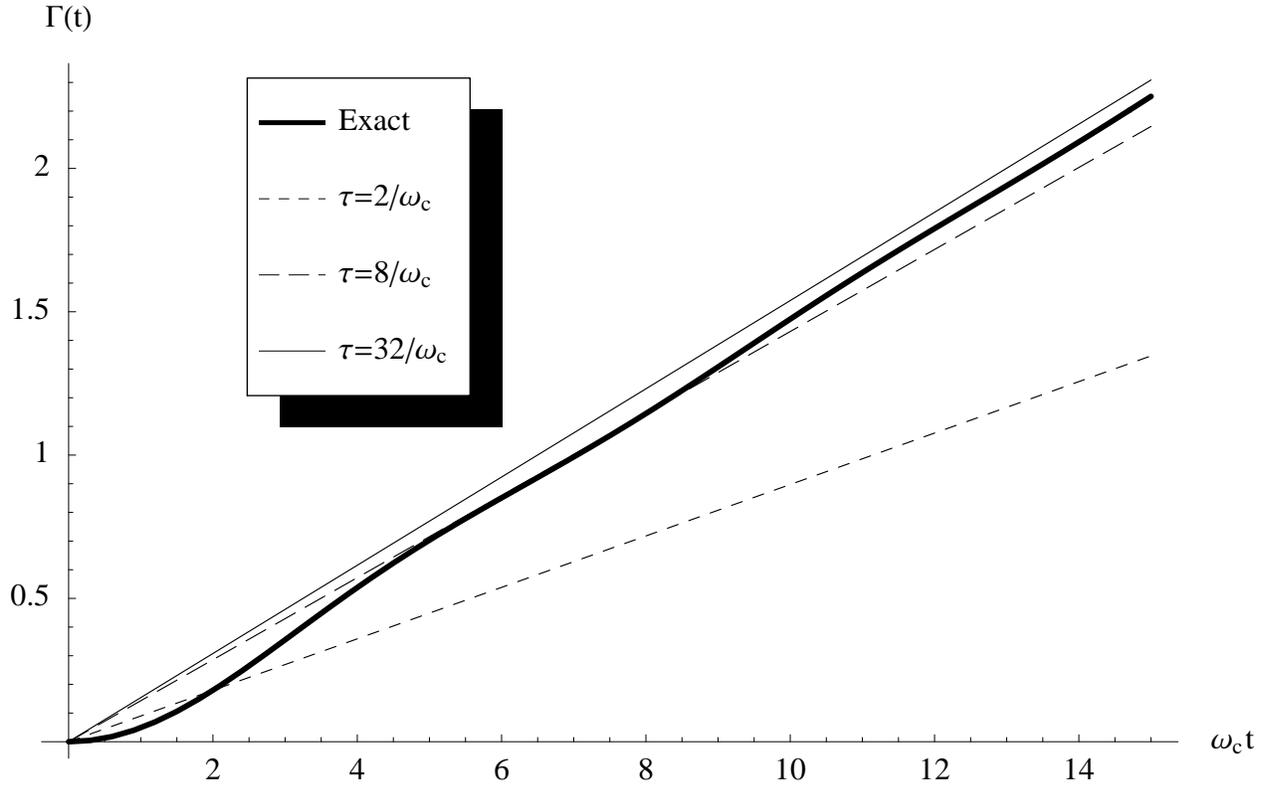,width=1.\textwidth}
\vspace{0.5cm}
\caption{Comparison of exact solution of the spin-boson model for
single-qubit pure dephasing to the result obtained from the Markovian master 
equation. Straight
lines correspond to the Markovian solution, which intersects the exact
solution (thick line) at $t=\tau$, as seen from
Eqs.~(\ref{eq:SME-Debye}),(\ref{eq:exact-Debye}). The density of states
of the boson bath is represented by the Debye model here.  Data is plotted for
$C=0.05$ and $\omega_c=1$.}
\label{figure}
\end{figure}

\end{document}